\documentclass[aps,amsmath,amssymb,showpacs,showkeys]{revtex4} %,twocolumn
\usepackage[dvips]{graphicx,color}
\usepackage{times}
\usepackage{multirow}
\usepackage{xcolor}
\usepackage[
  colorlinks=true,
  urlcolor=magenta,
  linkcolor=red,
  citecolor=blue]{hyperref}
\newcommand{\orcid}[1]{\href{https://orcid.org/#1}{\includegraphics[width=8pt]{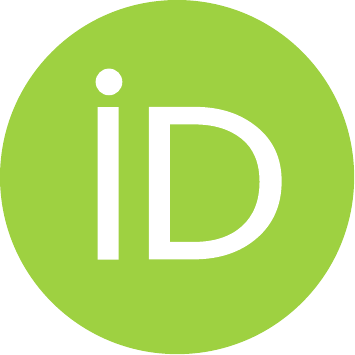}}}

\DeclareFontFamily{OT1}{pzc}{}
\DeclareFontShape{OT1}{pzc}{m}{it}{<-> s * [1.200] pzcmi7t}{}
\DeclareMathAlphabet{\mathpzc}{OT1}{pzc}{m}{it}

\begin{document}
\title{Strange stars in $f(\mathcal{R})$ gravity Palatini formalism and 
gravitational wave echoes from them}

\author{Jyatsnasree Bora \orcid{0000-0001-9751-5614}}
\email[Email: ]{jyatnasree.borah@gmail.com}

\author{Dhruba Jyoti Gogoi \orcid{0000-0002-4776-8506}}
\email[Email: ]{moloydhruba@yahoo.in}

\affiliation{Department of Physics, Dibrugarh University,
Dibrugarh 786004, Assam, India}

\author{Umananda Dev Goswami \orcid{0000-0003-0012-7549}}
\email[Email: ]{umananda2@gmail.com}

\affiliation{Department of Physics, Dibrugarh University,
Dibrugarh 786004, Assam, India}

%\date{}
\begin{abstract}
The compact stars are promising candidates associated with the generation of 
gravitational waves (GWs). In this work, we study a special type of compact 
stars known as strange stars in the $f(\mathcal{R})$ gravity Palatini 
formalism. Here we consider three promising $f(\mathcal{R})$ gravity models 
viz., Starobinsky, Hu-Sawicki and Gogoi-Goswami models in the domain of MIT 
Bag model and linear equations of state (EoSs). We compute the stellar 
structures numerically and constrained the $f(\mathcal{R})$ model parameters 
with a set of probable strange star candidates. The study shows that the 
consideration of stiffer MIT Bag model and linear EoSs within a favourable 
set of $f(\mathcal{R})$ gravity model parameters may result in strange stars 
with sufficient compactness to produce echoes of GWs. Thus, we have computed 
the GWs echo frequencies and characteristic echo times for such stars. It is 
found that in compliance with the experimentally obtained possible strange star 
candidates, the obtained GW echo frequencies for all the models are in the 
range of $65 - 85$ kHz. 
 
\end{abstract}

\pacs{}
\keywords{Strange Star; Equation of states; Modified theories of gravity; Gravitational wave echo.}

\maketitle

%%%%%%%%%%%%%%%%%%%%%%%%%%%%%%%%%%%%%%%%%%%%%%%%%%%%%%%%%%%%%%%%%%%%%%%%%%%%%%%%%%%%%

\section{Introduction}\label{intro}
From the current astrophysical and cosmological points of view, we are living 
in the era of accelerated expansion of the universe \cite{Spergel_2007,Astier_2006} 
as well as in the mystery of missing mass in the universe \cite{nashelski}. 
There are two general hypotheses to explain this present accelerated expansion 
of the universe. According to one hypothesis, Einstein's theory of general 
relativity (GR) requires the existence of some exotic energy, called the dark 
energy to explain this current state of the universe \cite{Riess_1998,perl}. 
In contrast, as per the second hypothesis the need for such an exotic energy 
can be avoided by extending the geometric part of the GR, usually referred to 
as the modified theories of gravity (MTGs), for the reason that it is feasible 
to obtain the present accelerated universe without the necessity of such an 
illusive exotic energy (detailed review can be found in \cite{cappo}). 
Similarly, the enigma of the missing mass of the universe has been explained by
introducing a non-luminous and non-baryonic unknown form of matter, dubbed as 
the dark matter \cite{berton}. However, from the perspective of the MTGs this 
missing mass can be explained without the need of the idea of dark matter 
\cite{Bohmer, Parbin}. Besides these, there are well-known observational and 
theoretical facts that compel us to study MTGs \cite{harko, nojiri}. In fact, 
currently MTGs are viewed hopefully as a good approach to go beyond 
GR \cite{clifton, nojiri2}.

To point out new physics from MTGs under the extreme relativistic effects, 
which are harder to achieve from the earth-based experiments, the compact stars 
give an excellent platform to do so. It is because of their unique high matter 
densities than the other nuclear matters. The family of compact objects mainly 
includes neutron stars and white dwarfs \cite{glendeinng, shapiro}. Such 
objects are the remnant of luminous stars, i.e.\ the endpoint of their stellar 
evolution. The family of such stars is now extended to include some more 
hypothetical stars, such as the strange stars \cite{aclock}, hybrid stars 
\cite{alford}, gravastars \cite{visser}, boson stars \cite{jetzer}, axion 
stars \cite{braaten}, and other compact exotic stars 
[e.g.~see \cite{dejan}]. Compared to other compact stars, by assumption, 
strange stars have a stable configuration and as such, perhaps they are 
composed of a matter that may be the true ground state of the hadronic 
matter \cite{witten,farhi}. Due to their unique mass-radius (M-R) relations, 
they are gaining the utmost interest in recent days. 
In white dwarfs and neutron stars, the degeneracy pressures are due to 
electrons and neutrons respectively. This degeneracy pressure is balanced by 
gravitational force to give a stable configuration to the star. If neutron 
stars get squeezed at a high temperature, decomposition of neutrons into 
component quarks occurs \cite{baym}. Such stars made of quark matter or 
de-confined quarks are known as the strange stars \cite{alcock, haensel}. Since 
they are in a much more stable configuration in comparison with neutron stars, 
so the proper understanding of these stars may lead to explain the origin of 
the huge amount of the energy released in super-luminous supernovae, which 
occur in about one out of every 1000 supernovae explosions, and they are 100 
times more luminous than regular supernovae \cite{ofek,ouyed}.

Literature surveys show that various authors have been studying the strange 
stars and their stellar structures in the realm of GR \cite{overgard, lopeZ, 
Maurya, jb2}. The properties of such compact stars have also been studied in 
the framework of MTGs by a number of authors \cite{maurya, pano, stay, deb, 
pretel, bhar, biswas, rej, hendi, panah, asta}. Such studies of compact stars 
in MTGs might allow one to impose constraints on the parameters responsible 
for the strong field regime, which are expected to break down in GR 
consideration. It needs to mention that the $f(\mathcal{R})$ gravity 
models in Palatini formalism are found to pass the astrophysical tests like 
post-Newtonian (solar system), parametrized post-Newtonian tests, 
etc.~\cite{olmo05, Olmo}. These theories are compatible with the current 
gravitational wave (GW) observations having good phenomenology in black holes 
\cite{11olmo, 12olmo, bejarano}, wormholes \cite{bambi, 16olmo}, and stellar 
structures \cite{olmo19, kainu, panotopoulos}. An important aspect of application 
of $f(\mathcal{R})$ gravity models that are designed to describe a low curvature 
phenomenon to the field of compact stellar study is that such models permit the 
presence of more massive compact stellar objects as compared to standard GR 
\cite{harko, goswami}. GR gives the maximal mass limit of compact star as 
$2\, M_{\odot}$ \cite{berti, bhatti, 20olmo}, while there are some more massive 
stellar structures exceeding the GR limit \cite{nice, van, rhoa, kalo, alsi}. 
Another motivation behind choosing $f(\mathcal{R})$ gravity models in high 
curvature regimes is that although the data available from the strong curvature 
regime phenomena are compatible with the GR, there are some issues related to such 
data which seem to complying with the $f(\mathcal{R})$ and other MTGs 
\cite{kono, vainio}. Moreover, at the astrophysical scales there are some 
persistent issues of observed massive pulsars which can hardly be explained 
within the GR limit \cite{grigorian, lenzi, lihu}. So a possible way for getting 
over these observational issues are MTGs. Interestingly compact objects like 
neutron stars and strange stars are indeed good tools to test and/or constrain 
MTGs. Thus it can be conjectured that the strong gravity regimes are some 
important tools to check the viability of different MTG models \cite{psaltis}.
Previously, compact stars in $f(\mathcal{R})$ theories of gravity in the 
Palatini formalism have been investigated in Refs.\ \cite{kainu, olmo, pannia, 
panotopoulos, Silveira}. In Ref.\ \cite{panotopoulos}, G.\ Panotopoulos 
studied the structure of strange stars by considering two models of 
$f(\mathcal{R})$ theories viz., the Starobinsky model: 
$f(\mathcal{R})=\mathcal{R}+\mathcal{R}^{2}/(6\mathcal{M}^{2})$ 
and the $1/\mathcal{R}$ model with a constant trace $T$ of the 
energy-momentum tensor $T_{\mu\nu}$. He showed that the behaviour of strange 
stars varies depending on the model type \cite{panotopoulos}. Recently, 
Silveira et.\ al.\ studied \cite{Silveira} the compact and ultra-compact
objects considering a Schwarzschild type homogeneous star and by adding a transition 
zone for the density near the surface of star in the $f(\mathcal{R})$-Palatini 
theory. In GR the asteroseismological behaviours of strange stars are studied in 
\cite{jb} and also the GW echo frequencies emitted by such 
stars are reported earlier in \cite{pani, manarelli, urbano, jb, zhang}. Motivated 
by these works, in this present work we have used constant as well as variable 
trace $T$ of the energy-momentum tensor $T_{\mu\nu}$ to see how the structure 
of such stars depends on models of the $f(\mathcal{R})$ gravity in Palatini 
formalism. We have considered three concrete and viable $f(\mathcal{R})$ 
gravity models i.e., the Starobinsky model \cite{staro}, the Hu-Sawicki model 
\cite{hu}, and the newly introduced model in \cite{gg}. It needs to be 
mentioned that we consider this new model with the other two well-known 
models in the field of MTGs to validate the new model in the astrophysical 
context. Hereafter, for the convenient representation we will call this model 
as the Gogoi-Goswami model. 
Using the modified Tolman-Oppenheimer-Volkoff (TOV) equations for the 
$f(\mathcal{R})$ gravity along with three equations of state (EoSs) viz., 
the MIT Bag model EoS, the stiffer MIT Bag model EoS and 
the linear EoS, we found the stellar structure solutions. We have 
constrained these models using a set of possible strange stars candidates. 
Also, for the first time in this work we have determined the GW echo 
frequencies emitted by these ultra-compact objects in the $f(\mathcal R)$ 
gravity Palatini formalism. 

After this brief introduction, we have organized our article as follows. In 
section \ref{mtg}, the $f(\mathcal{R})$ gravity theory in Palatini formalism 
and the concerned models of this gravity are discussed very shortly. The 
modified TOV equations are derived and the method of their corresponding 
solutions for the considered models are also discussed in this section. A 
workable introduction of GW echoes is given in the section \ref{gwe}. In 
section \ref{numerical}, the numerical results of our work are presented. 
Finally, we conclude our article in the section \ref{conclusion}. In this 
article, we adopt the unit of $c=G=1$, where $c$ and $G$ denote the speed of 
light and the gravitational constant respectively, and also we used the 
metric signature $(-,+,+,+)$.
 
%%%%%%%%%%%%%%%%%%%%%%%%%%%%%%%%%%%%%%%%%%%%%%%%%%%%%%%%%%%%%%%%%%%%%%%%%%%%%%%%%%%%%

\section{Ultracompact stars in modified theories of gravity}\label{mtg}
In this section, first we shall briefly summarize the $f(\mathcal{R})$ theories 
of gravity in the Palatini formalism and then shall derive the modified TOV 
equations in this theory. These TOV equations are then numerically solved for
the three viable $f(\mathcal{R})$ gravity models using the three EoSs as 
mentioned in the previous section to study the properties of strange stars.

In Palatini $f(\mathcal{R})$ theories the action is given by
\begin{equation}
	\label{j}
	S=\dfrac{1}{16\pi}\int d^{4}x\,\sqrt{-g}f(\mathcal{R})+\int d^{4}x\sqrt{-g}\,\mathcal{L}_{m}(g_{\mu\nu}),
	\end{equation}
where $\mathcal{L}_{m}$ is the matter Lagrangian density and $f(\mathcal{R})$ 
is the generic function of the Ricci scalar. Varying this action with respect 
to both the metric $g_{\mu\nu}$ and the affine connection 
$\Gamma^{\rho}_{\mu\nu}$, the field equations can be obtained in the 
form \cite{kainu}:
\begin{equation}
	\label{k}
	F(\mathcal{R})\mathcal{R}_{\mu\nu}-\dfrac{1}{2}f(\mathcal{R})g_{\mu\nu}=8\pi T_{\mu
	\nu},
	\end{equation}
\begin{equation}
	\label{l}
	\nabla_{\rho}\!\left(\sqrt{-g}F(\mathcal{R})g^{\mu\nu}\right)=0.
	\end{equation}
Here $F\equiv\partial f/\partial \mathcal{R}$, $\mathcal{R}\equiv g^{\mu\nu}\mathcal{R}_{\mu\nu}(\Gamma)$ and $T_{\mu\nu}$ is the energy-momentum tensor of $\mathcal{L}_{m}$. 
Taking the trace of equation \eqref{k}, a purely algebraic 
equation relating $\mathcal{R}$ and $T$ can be obtained as
\begin{equation}
	\label{m}
	F(\mathcal{R})\mathcal{R}-2f(\mathcal{R})=8\pi T.
	\end{equation}

If we consider the perfect fluid model for the materials of stars, the 
trace of $T_{\mu\nu}$ can be obtained as $T=-\,\rho + 3p$ with the pressure 
$p$ and the energy density $\rho$. Again, in order to describe a spherically 
symmetric spacetime, we use the metric represented by the line element:
\begin{equation}
        \label{b}
        ds^{2} = -f(r)\,dt^{2} + g(r)\,dr^{2} + r^{2}d\Omega^{2},
    \end{equation}
where the unknown metric functions $f(r)$ and $g(r)$ are dependent on the 
radial coordinate $r$ only. $d\Omega^{2}=d\theta^{2}+\sin^{2}\theta\, 
d\phi^{2}$ with $\theta$ and $\phi$ are the polar angle and the azimuth angle 
respectively. Considering the exterior to the stars be the vacuum with 
$T_{\mu\nu}=0$, the exterior solutions for our considered models are found 
to be the Schwarzchild (de-Sitter if $\Lambda_0>0$ or anti-de-Sitter if 
$\Lambda_0<0$) solutions as given by 
\begin{equation}
	\label{c}
	f(r)=g(r)^{-1}=1-\dfrac{2M}{r}{ -\dfrac{\Lambda_0}{3}r^2}.
	\end{equation}
Whereas the interior solutions in this theory give,
\begin{equation}
	\label{d}
	f(r) = g(r)^{-1}=1-\dfrac{2m(r)}{r}-\dfrac{\Lambda_\rho}{3}r^2.
	\end{equation}
Here $M$ is the total mass of stars with the radius $R$ and $m(r)$ is their 
mass at the radial distance $r<R$. $\Lambda_{\rho}$ is the effective 
cosmological constant inside the stars, which arises due to the curvature 
function $f(\mathcal{R})$ of the theory and is given by
\begin{equation}
        \label{t}
        \Lambda_{\rho}=\dfrac{1}{2}\left(\mathcal{R}-\dfrac{f(\mathcal{R})}{F(\mathcal{R})}
        \right).
        \end{equation}
Upon matching the exterior and interior solutions at $r=R$, we can obtain the total 
mass of stars as \cite{kainu,panotopoulos}
\begin{equation}
	\label{a}
	M = m(R)+\dfrac{(\Lambda_\rho{-\Lambda_0})R^3}{6}.
	\end{equation}
It is to be noted that the vacuum value of the effective cosmological 
constant $\Lambda_0$ outside the star vanishes for our considered models. 
Thus, for our models equation \eqref{a} will take the form:
\begin{equation}
M = m(R)+\dfrac{\Lambda_\rho R^3}{6}.
\end{equation} 
One can see that the curvature modification has a contribution to the total mass
of stars in the theory depending on the type of a model. The bare mass $m(R)$ 
in this mass equation can be calculated from the modified TOV equations given 
below. Now, to be more specific we consider $g(r) = e^{\lambda (r)}$ and $f(r) = e^{\chi (r)}$, where $\lambda(r)$ and $\chi(r)$ are two source 
functions of $r$. For the line element \eqref{b} with these specific forms of 
$f(r)$ and $g(r)$, the tt- and rr-components of field equations \eqref{k} and 
\eqref{l} give the equations for functions $\lambda(r)$ and $\chi(r)$ as
\begin{align}
	\label{ad1}
	\lambda'(r)&=\dfrac{1}{1+\gamma}\left(\dfrac{1-e^{\lambda}}{r}+\dfrac{\alpha+\beta}{r}+8\pi r\rho \,\dfrac{e^{\lambda}}{F}\right),\\[5pt]
    \label{ad2}
	\chi'(r)&=-\dfrac{1}{1+\gamma}\left(\dfrac{1-e^{\lambda}}{r}+\dfrac{\alpha}{r}-8\pi r p\, \dfrac{e^{\lambda}}{F}\right),
	\end{align}
where the prime denotes the derivative with respect to $r$ and
\begin{equation}
	\label{n}
	\alpha=r^{2}\left[\dfrac{4}{3}\left(\dfrac{F'}{F}\right)^{2}+\dfrac{2}{r}\dfrac{F'}{F}+\dfrac{e^{\lambda}}{2}\left(\mathcal{R}-\dfrac{f}{F}\right)\right],\;\;
	\beta=r^{2}\left[\dfrac{F''}{F}-\dfrac{3}{2}\left(\dfrac{F'}{F}\right)^{2}\right],\;\;\mbox{and}\;\;
	\gamma=\dfrac{rF'}{2F}.
	\end{equation}

For a star in hydrostatic equilibrium the interior solution is described by 
the famous TOV equations \cite{tolman,tov}. The solutions of these equations 
can lead one to know the physical properties like mass, pressure and radius of
the star. These TOV equations will take the modified form in MTGs in comparison
to their form in GR. The modified TOV equations for the $f(\mathcal{R})$ 
gravity in the Palatini formalism are found as \cite{kainu}
\begin{align}
	\label{q}
	\dfrac{dp}{dr}&=-\dfrac{1}{1+\gamma}\dfrac{p+\rho}{r(r-2m)}\left[m+\dfrac{4\pi r^{3} p}{F}-\dfrac{\alpha}{2}(r-2m)\right],\\[10pt]
	\label{r}
	\dfrac{dm}{dr}&=\dfrac{1}{1+\gamma}\left[\dfrac{4\pi r^{2}\rho}{F}+\dfrac{\alpha+\beta}{2}\dfrac{m}{r}(\alpha+\beta-\gamma)\right].
	\end{align}
%In these equations in the right hand sides $F'$ and $F''$ are first and 
%second ordder derivative of pressure, e.g., $F'=d/dr[F(\mathcal{R}(T))]$.
It is to be noted that when $F=1$, then these equations will take their 
original form in GR. Moreover, when the trace of the energy-momentum tensor 
$T(\rho)$ is a constant independent of $r$, then
% density $\rho$ is a constant independent of $r$, then $T\equiv T(\rho)$ and 
the Ricci scalar $\mathcal{R}$ is also a constant. One should note that 
the complete field equations equations \eqref{k} and 
\eqref{l} take the familiar form of Einstein's equations with a non-vanishing 
cosmological constant and with the right side scaled by a $\rho$ dependent 
constant parameter $F_{\rho}$ \cite{kainu} as
\begin{equation}
	\label{s}
	G_{\mu \nu} +\Lambda_{\rho} g_{\mu\nu} = \dfrac{8\pi}{F_{\rho}}\, T_{\mu\nu}.
	\end{equation}
Here $F_{\rho}$ is the derivative of $f(\mathcal{R})$ with respect to 
$\mathcal{R}$ evaluated at the constant value $\mathcal{R}_0$ that solves the 
trace equation (equation \ref{m}). $\Lambda_{\rho}$ is the effective 
cosmological constant as given in equation  \eqref{t} and for a constant trace 
$T$ it is independent of $r$.

%However, for variable trace $T(r)$ complete field equations Eqn.\,\ref{q} and 
%\ref{r} are to be integrated. 
%Now, in $f(R)$ theories of gravity there are two different cosmological constants inside and outside the star. Upon matching these two solutions one can obtain the modified mass term unlike in GR \cite{pano, kainu} as \begin{equation} \label{s} M=m(R)+\dfrac{\Lambda_{\rho}-\Lambda_{0}}{6}\mathcal{R}^{3}\end{equation} When one applies the EoS corresponding to a relativistic gas of deconfined quarks, for the exterior solution the energy-momentum vanishes, (since in vacuum $T_{\mu\nu}=0$) and the effective cosmological constant becomes $\Lambda_{\rho}=\Lambda_{0}$ at $T=0$. Using TOV equation together with EoSs and matching conditions,the problem is reduced to standard GR problem with a non-vanishing cosmological constant. But the pressure and energy density for the matter fluid are rescaled by a factor $1/F_{\rho}$.

At this moment we are in a position to use the $f(\mathcal{R})$ gravity models 
of our interest as mentioned earlier. The Starobinsky model with its extended 
form can be written as \cite{staro}
\begin{equation}
	\label{u}
	f_{s}(\mathcal{R})=\mathcal{R}-s\mathcal{M}_{s}^{2}\left[1-{\left(\dfrac{\mathcal{R}^{2}}{\mathcal{M}_{s}^{4}}+1\right)}^{-n}\right],
	\end{equation}
where $s$, $\mathcal{M}_s$ and $n$ are the Starobinsky model parameters. In 
fact, the parameter $\mathcal{M}_{s}$ is the mass scale and here we consider 
it as a free parameter. In our work we will use $n=1$, for which the model 
reduces to the form:
\begin{equation}
	\label{v}
	f_{s}(\mathcal{R})=\mathcal{R}-s\,\dfrac{\mathcal{R}^{2}}{\mathcal{M}_{s}^{2}}.
	\end{equation}
The first derivative of the above equation with respect to $\mathcal{R}$ gives,
\begin{equation}
	\label{w}
	F_{s}(\mathcal{R})=1-2s\dfrac{\mathcal{R}}{\mathcal{M}_{s}^{2}}.
	\end{equation}
Using this result with the model \eqref{v} in the trace equation \eqref{m}, we 
can obtain the background curvature
\begin{equation}
	\label{x}
	\mathcal{R}_{0}=-\,8\pi T.
	\end{equation}

The second $f(\mathcal{R})$ gravity model considered in this work is the 
Hu-Sawicki model, which is defined as \cite{hu}
\begin{equation}\label{y}
f_h (\mathcal{R}) = \mathcal{R} - \dfrac{c_1 \mathcal{M}^2_h \left(\dfrac{\mathcal{R}}{\mathcal{M}^2_h}\right)^\mu}{c_2 \left(\dfrac{\mathcal{R}}{\mathcal{M}^2_h}\right)^\mu + 1},
\end{equation}
where $c_1$, $c_2$ and $\mathcal{M}_h$ are model parameters. The parameters 
$c_1$ and $c_2$ are dimensionless in nature, while the third parameter 
$\mathcal{M}_h$ has the dimension of length$^{-1}$, which is considered to be 
a free parameter here.
%\begin{equation}	\label{y}
%		f_h(\mathcal{R})= \mathcal{R}-\dfrac{c_1\; \mathcal{M}_h^2
%	\left(\dfrac{\mathcal{R}}{\mathcal{M}_h^2\;\right)^\mu} }{c_2
%		\left(\dfrac{\mathcal{R}}{\mathcal{M}_h^2}\right)^\mu+1}
%	\end{equation}
For this model we have,
\begin{equation}\label{z}
F_h (\mathcal{R}) = 1+ \dfrac{c_1 c_2 \mu \left(\dfrac{\mathcal{R}}{\mathcal{M}^2_h}\right)^{2\mu-1}}{\left[ 1+ c_2 \left( \dfrac{\mathcal{R}}{\mathcal{M}^2_h} \right)^\mu \right]^2}.
\end{equation}
 
% \begin{equation}
%	F_h(\mathcal{R})= 1+\dfrac{c_{1}c_{2}\mu\left(\frac{\mathcal{R}}{\mathcal{M}_{h}^{2}\right)^{2\mu-1}}}{\left(1+c_{2}\left(\frac{\mathcal{R}}{\mathcal{M}_{h}^2}
%	\right)^{\mu}\right)^2}
%\end{equation}
From the trace equation \eqref{m}, we can find out the background curvature for 
this model as given by 
\begin{equation}
	\label{aa}
         \mathcal{R}_{0} = \frac{A^{2/3} + B}{3\,c_2\,C^{1/3}},
	%\mathcal{R}_{0}= \frac{B^{2/3}+2 \sqrt[3]{B} \left((c_1 -1) \mathcal{M}_{h}^2 -4 \pi c_2 T\right)-16\pi (2c_1 +1) c_2 \mathcal{M}_{h}^2 T+(c_1 -1) (4 c_1 -1) \mathcal{M}_{h}^4 + 64 \pi ^2 c_1 ^2 T^2}{3 c_2 \sqrt[3]{A+12 \pi \left(-8 c_1 ^2+c_1 -2\right) c_2 \mathcal{M}_{h}^4 T+192 \pi ^2 (2 c_1 +1) c_2 ^2 \mathcal{M}_{h}^2 T^2+(c_1 -1)^2 (8 c_1 +1) \mathcal{M}_{h}^6-512 \pi ^3 c_2 ^3 T^3}},
\end{equation}
where 
	\begin{align}
	A &= D+192 \pi ^2 (2 c_1 +1) c_2 ^2 \mathcal{M}_{h}^2 T^2-12 \pi  (c_1 (8 c_1 -1)+2) c_2 \mathcal{M}_{h}^4 T+(c_1 -1)^2 (8 c_1 +1) \mathcal{M}_{h}^6-512 \pi ^3 c_2 ^3,\\[5pt]
        B & = 2\,C^{1/3} \left[(c_1 -1) \mathcal{M}_{h}^2 -4 \pi c_2 T\right]-16\pi (2c_1 +1) c_2 \mathcal{M}_{h}^2 T+(c_1 -1) (4 c_1 -1) \mathcal{M}_{h}^4 + 64 \pi ^2 c_1 ^2 T^2,\\[5pt]
        C & = D+12 \pi \left(-8 c_1 ^2+c_1 -2\right) c_2 \mathcal{M}_{h}^4 T+192 \pi ^2 (2 c_1 +1) c_2 ^2 \mathcal{M}_{h}^2 T^2+(c_1 -1)^2 (8 c_1 +1) \mathcal{M}_{h}^6-512 \pi ^3 c_2 ^3 T^3.\\[5pt]
        D &= 3 \mathcal{M}_{h}^3 \sqrt{-\,3\,c_1\big[-48 \pi ^2 (5 c_1 +4) c_2 ^2 \mathcal{M}_{h}^2 T^2+24 \pi  (c_1 -1)^2 c_2 \mathcal{M}_{h}^4 T+(c_1 -1)^3 \mathcal{M}_{h}^6+512 \pi ^3 c_2 ^3 T^3\big]}.
	\end{align}

The third viable $f(\mathcal{R})$ gravity model considered in this work is 
the Gogoi-Goswami model having the form \cite{gg}:
\begin{equation}
	\label{bb}
	f_g(\mathcal{R})=\mathcal{R}-\dfrac{\alpha}{\pi}\, 
	\mathcal{R}_c \cot^{-1}	
	\left(\dfrac{\mathcal{R}^2_c}{\mathcal{R}^2} \right)
	-\beta\,\mathcal{R}_c\left[1-	
	\exp \left(-\frac{\mathcal{R}}{\mathcal{R}_c}\right)\right],
	\end{equation}
where $\alpha$ and $\beta$ are two free parameters of the model, and 
$\mathcal{R}_c$ is the characteristic curvature constant with the dimension 
same as the curvature scalar $\mathcal{R}$. The first derivative of the model 
with respect to $\mathcal{R}$ is given by
\begin{equation}
	\label{cc}
	F_{g}(\mathcal{R})=1-\frac{2\alpha \mathcal{R}\mathcal{R}_{c}^3}{\pi\mathcal{R}^4+
	\pi\mathcal{R}_{c}^4}-\beta e^{-\frac{\mathcal{R}}{\mathcal{R}_{c}}}.
	\end{equation}
In case of this model, the trace equation \eqref{m} takes the following form:
\begin{equation}
	\label{dd}
	\mathcal{R}_c \left[x \left(-\frac{2\, \alpha\,  x}{\pi  x^4+\pi }-\beta  e^{-x}-1\right)+\frac{2 \alpha  \tan ^{-1}(x^2)}{\pi }+2 \beta\left(\sinh (x)-\cosh (x)+1\right)\right]=8 \pi  T,
	\end{equation}
where $x = \mathcal{R}/\mathcal{R}_c$ is a dimensionless variable. This 
equation can't be solved analytically and hence we have to implement a 
numerical method to solve the same. For this purpose, we define two functions 
as
\begin{align}
f_{g}(x)& = x \left(-\frac{2 \alpha  x}{\pi  x^4+\pi }-\beta  e^{-x}-1\right)+\frac{2 \alpha  \tan ^{-1}(x^2)}{\pi }, \\[5pt]
 g_{g}(x)& =  \frac{8 \pi  T}{\mathcal{R}_c}-2 \beta\, \left[\sinh (x)-\cosh (x)+1\right],
\end{align}
where $x$ can be a solution of equation \eqref{dd} provided $$\big| f_g(x) - g_g(x)\big| \approx 0.$$

Further, to describe the ultra compact stars, in particular strange stars we 
shall consider three EoSs. The first one is the MIT Bag model EoS, which was
proposed to explain a relativistic gas of massless de-confined quarks 
\cite{witten,chodosa,chodosb, jb} and is given by
\begin{equation}
	\label{ee}
	p = \dfrac{1}{3} (\rho - 4B). 
	\end{equation}
Here $p$ is the perfect fluid pressure, $\rho$ is the fluid density and $B$ is 
the Bag constant. The constant $B$ may have different possible values, which
have been discussed in detail in Ref.\ \cite{jb}. In this work we shall use 
a suitable value of $B=(168MeV)^{4}$. This value is chosen in such 
a way that it lies in the acceptable range \cite{aziz} and the solutions 
obtained with this value are such that they do not exceed the mass and radius 
of $\approx3\,\mbox{M}_\odot$ and $\approx 13\,\mbox{km}$ respectively for 
all the considered cases in the present study. This mass and radius bound 
is taken in such a way that we will get physically stable strange star configurations in our study.
However, to obtain the stellar properties, especially the property that is suitable to 
get GW echoes from the stars, we have also chosen the stiffer form of this EoS 
\cite{manarelli, jb} as given by
\begin{equation}
	\label{ff}
	p = (\rho - 4B). 
	\end{equation}
The third EoS considered in this work is the linear EoS \cite{dey} having the 
form:
\begin{equation}
	\label{gg}
	p = b(\rho - \rho_{s}), 
	\end{equation}
where $b$ is the linear constant and $\rho_{s}$ is the surface energy 
density of the star. As in the case of the MIT Bag model EoS, for this EoS also
we have taken a suitable value of $b = 0.910$ \cite{jb} in this work. 
By fixing some EoSs the TOV equations (in GR or in MTGs) can be solved 
numerically, which will lead to know the structure of the compact object under 
consideration. Thus with these three EoSs we have solved numerically the TOV 
equations in the $f(\mathcal{R)}$ gravity models of Starobinsky, Hu-Sawicki and 
Gogoi-Goswami in the Palatini formalism. Using the solutions of these stellar 
structure equations, finally we have calculated the GW echo frequencies emitted
by such stars.

From an EoS we can have a specific form of the energy-momentum trace, which 
can be used in the expression of the background curvature for each 
$f(\mathcal{R})$ gravity model to obtain a specific form for its background 
curvature. In the case of the MIT Bag model (equation \eqref{ee}), the trace 
of the energy-momentum $T$ is a constant quantity leading to a simple 
situation. While for the stiffer form of the MIT Bag model EoS 
(equation \eqref{ff}) the energy-momentum trace $T= 2\,\rho -12B$ and for 
the linear EoS (equation \eqref{gg}) it is $T=(3\,b-1)\rho -3\rho_s$. Thus, 
using these EoSs in the modified TOV equations \eqref{q} and \eqref{r}, and 
also using the considered $f(\mathcal{R)}$ gravity models with the initial 
conditions: $m(r=0)=0$ and $p(r=0)=p_{c}$, we can solve them numerically. In 
addition the bare mass $m(R)$ and the radius $R$ of a star can be determined 
by the boundary conditions: $m(r = R) = m(R)$ and $p(r = R) = 0$. 

One should note that the gravitational field for strange stars is sourced by 
the trace of the energy-momentum tensor as given by equation \eqref{m}. In 
the usual case, for the Sun like stars the pressure to density ratio is very 
small ($\approx 10^{-9}$) and so it is possible to neglect the pressure for 
non-relativistic stars while calculating the metric. But in the case of stars 
with high compactness such as strange stars, we can not neglect the pressure 
term, which increases the complexity of the problem. However, in the case of 
the standard MIT Bag model, we have $T=constant$, which makes the situation 
less complex. One may note here that $T=constant$ does not imply that the 
density of a star is constant throughout the whole region, but instead it 
indicates that any variation in the pressure and density of the star does not 
influence the trace of the energy-momentum tensor. So, undoubtedly this is an 
ideal standard case in the $f(\mathcal{R})$ gravity, which makes the situation 
comparatively simple enough to handle. The complexity increases when we move 
to the other EoSs, where the trace of the energy-momentum tensor is not 
constant, i.e.\ any variation in the pressure and density of a star can impact 
the trace of the energy-momentum tensor significantly. We have seen that a
viability condition $\big| F(\mathcal{R})-1 \big| \ll 1$ makes the system 
solvable under suitable approximations. Another thing to consider is the 
matching of exterior and interior solutions i.e., while solving field 
equations of $f(\mathcal{R})$ gravity, junction conditions must be considered 
for the smooth matching of interior and exterior spacetimes \cite{Olmo}. In GR, 
these boundary conditions are known as the Darmois-Israel matching or junction 
conditions \cite{israel,darmois}. In the Ref.\ \cite{Olmo} the authors 
utilized the tensor distributional technique to find the junction 
conditions for Palatini $f(\mathcal{R})$ gravity. It is shown that these 
junction conditions are required to build the stellar models of gravitational 
bodies with the matching of interior and exterior regions at some 
hypersurfaces. 
They have considered the stellar surfaces in the polytropic models in the 
Palatini framework as an example and their results showed that the white dwarfs 
and neutron stars can be modelled safely within this framework. For the smooth 
matching of our exterior and interior solutions on the stellar surface, we 
consider a hypersurface $\Sigma$ which separates the interior stellar 
structure from the exterior vacuum. At the junction $r=R$, the continuity of 
the geometry can be ensured by the condition \cite{shariff, pk},
\begin{equation}\label{junc1}
ds^2_{-}\vline_{\,\Sigma}=ds^2_{+}\vline_{\,\Sigma}
\end{equation}
with $-$ indicating the interior part and $+$ denoting the exterior one.
This junction condition will result the equality:
\begin{equation}
1-\dfrac{2m(R)}{R}-\dfrac{\Lambda_\rho R^2}{3}=1-\dfrac{2M}{R}-\dfrac{\Lambda_0 R^2}{3}.
\end{equation}
From this equality condition the total mass of the stellar structure can be 
retrieved as given in equation \eqref{a}. From this condition it can be 
directly followed that the trace $T$ of the energy momentum tensor 
$T_{\mu \nu}$ must be continuous across the surface $\Sigma$ and hence 
\begin{equation}\label{cont01}
\left[ T \right] \equiv T^+ - T^- = 0.
\end{equation}
This condition of the trace was already expected by the virtue of the 
continuity and standard differentiability of tensorial equations \cite{Olmo}. 
Respecting this condition the singular part of the field equations \eqref{k} 
can be expressed as \cite{Olmo}
\begin{equation}\label{cont02}
F_{\Sigma}\,\mathcal{G}_{\mu\nu} + F_{\mathcal{R}} \mathcal{R}_T\; \vline_{\,\Sigma}\, \mathpzc{b}\, h_{\mu\nu} = 8\pi \tau_{\mu\nu},
\end{equation}
where $\mathcal{G}_{\mu\nu}$ is the singular part of the Einstein tensor 
$G_{\mu\nu}$ on the surface $\Sigma$, $h_{\mu\nu}$ represents the projector on 
$\Sigma$. $\mathpzc{b}$ is a scalar quantity defined as 
\begin{equation}
\mathpzc{b}\equiv n^{\alpha}\left[\nabla_{\alpha} T\right]
\end{equation}
and $n$ is the unit vector normal to $\Sigma$.

For the case of the MIT Bag model, $\rho = 3p + 4B$ and hence $\rho_p=3,\;
\rho_{pp} = 0$. Again, the trace of the energy momentum tensor is considered 
as $T = 3p-\rho$. Thus the scalar quantity $\mathpzc{b}$ can be evaluated
for this EoS as
\begin{align}
\mathpzc{b} &= n_+^r \delta_r T^+ - n_-^r \delta_r T^-\\
&\approx 0 + (3 - \rho_p) p_r = (3-3)p_r 
 = 0
\end{align}
Since, this is the exclusive property of the MIT Bag model, so for any 
$f(\mathcal{R})$ gravity model this condition will hold good and consequently 
equation \eqref{cont02} will be formally identical to its GR counterpart 
\cite{Olmo}. Thus, one can easily see that for the MIT Bag model in 
$f(\mathcal{R})$ gravity, the field equations are divergence free and 
continuous resulting in viable compact star solutions. In case of the stiffer 
MIT Bag model EoS,
\begin{align}
\mathpzc{b} &= n_+^r \delta_r T^+ - n_-^r \delta_r T^- = 2\,p_r
\end{align}
Now, at the surface of the star, from equation \eqref{q} we may obtain the
expression of $p_r$ as
\begin{equation}\label{add1} 
\frac{dp}{dr}\big|_R \equiv p_R = -\,\frac{\rho_S C}{R(1-2C)},
\end{equation}
where $\rho_S$ is the density at the surface (which is non-zero for
our considered EoSs) and $C= M/R$ is the compactness of a given star. From 
this equation \eqref{add1} it can be shown that $|p_R|$ is a non-zero
small finite quantity less than unity for all $f(\mathcal{R})$ gravity models
with the corresponding sets of parameters considered in this study. However, 
in general, it is seen that $p_R$ will be diverse only when $2C=1$ or
when $C= 0.5$. This limiting value of $C$ is greater than the GR Buchdahl
limit $C=0.44$. Hence, in general also there is no chance of the divergence 
of $\mathpzc{b}$ in this EoS. At this point it is also to be noted that
$2C>1$ or $C> 0.5$ is not allowed by the Buchdahl limit as well as by the 
fact that for such values of $C$, $p_R$ becomes positive, which leads an 
unphysical situation. Again, the models chosen in this study are dark energy 
$f(\mathcal{R})$ gravity models satisfying the solar system tests. Such models 
satisfy $F(\mathcal{R})-1 \ll 1$ as mentioned above. For the selected parameter
sets of our study, this condition is valid and $F_{\mathcal{R}} \approx 0$ 
without introducing any divergences. Hence, in equation \eqref{cont02}, 
$F_{\mathcal{R}}$ vanishes near the surface of the star making the expression 
identical to its GR counterpart. Finally for linear EoS, $\mathpzc{b}$ can
be found as
\begin{align}
\mathpzc{b} &= n_+^r \delta_r T^+ - n_-^r \delta_r T^- = (3-1/b)p_r.
\end{align}
Similar to the above case here also we can see that $\mathpzc{b}$ is free from 
divergences and has a small value near the surface of the star for a given 
set of parameters within our considered range. As well in general it is free 
from divergence in this EoS as in the previous case. Further, for the 
considered $f(\mathcal{R})$ models, $F_{\mathcal{R}} \approx 0$ holds good 
without introducing any divergence, making equation \eqref{cont02} identical 
to the GR counterpart in the hypersurface $\Sigma$.
%In Palatini formalism, however such solutions are obvious and followed directly from the trace equation. The relation between the exterior mass and the interior density can be given as equation \eqref{a}. This relation plays a very important role and matches the interior and exterior solutions \cite{kainu}. For example 
Moreover, if we consider 
a Sun like star, $F_{\rho}$ can not significantly differ from the unity inside it, 
otherwise this will led to significant deviations from the predictions of Solar 
physics. It is because, the local density determines the other thermodynamical 
observational parameters of the star and the theory should be consisted with 
the Solar system observations. This simply implies that in $f(\mathcal{R})$ 
gravity Palatini formalism, Sun like stars have very very small deviations
from the GR calculation and most of it comes from the effective cosmological 
constants inside and outside the star i.e.\ from $\Lambda_{\rho}$ and 
$\Lambda_0$ respectively. A recent study shows that consideration of an 
observed amount of dark energy $\Omega_\Lambda \approx 0.72$ imposes almost 
negligible changes in the mass of a star \cite{kainu}. In this 
present study, we have calculated the interior and exterior effective 
cosmological constants from the considered $f(\mathcal{R})$ gravity models and 
matched the interior and exterior solutions for the continuity. So, obviously 
from the nature of the viable models and the continuity condition, 
$\Lambda_\rho$ can't have a very large difference from $\Lambda_0$. This makes 
it comparatively less complicated to solve the TOV equations numerically under 
suitable approximations and with precision.  

%%%%%%%%%%%%%%%%%%%%%%%%%%%%%%%%%%%%%%%%%%%%%%%%%%%%%%%%%%%%%%%%%%%%%%%%%%%%%%%%%%%%

\section{Gravitational wave echoes}\label{gwe}
An important property associated with the compact stars is that some of them 
are promising candidates to echo GW falling on their surface. Those compact 
stars which possess a photon sphere \cite{claudel} at distance of $R=3\,M$ are the only 
eligible candidates to emit GW echoes \cite{pani}. When GWs emerging from some 
merging events fall on the surface of such compact stars, they get eventually 
trapped by the photon sphere and then echoes of GWs emerge. The important 
criteria to bear this photon sphere in a star is that the compactness 
$C=M/R$ of that star should be, $1/3<C<4/9$ \cite{urbano}. This upper limit on 
compactness is the Buchdahl's limit with the Buchdahl's radius 
$R_B=9/4\,M$ \cite{buchdahl}, which describes the maximum amount of mass 
that can exist in a sphere before it must undergo the gravitational collapse 
\cite{urbano}. One should note that this limit is valid for the GR cases
only. For the MTGs this limit needs some modifications. Such modification in
the Buchdahl's limit for $f(\mathcal{R})$ theory in the metric formalism was 
reported in the Ref.\;\cite{goswami}. A more generalized study on the 
modification of this limit was reported in Ref.\;\cite{burikham}, which is 
valid for a large class of generalized gravity models. They have reported a 
bound for the maximum possible mass of a compact stellar object in MTGs as 
\cite{burikham}
\begin{equation}\label{buchdahl}
C\leq\dfrac{1}{9}(4-\dfrac{3a}{2}).
\end{equation}
Here $a=4\,\pi\, p_{eff}(R)R^2$ and $p_{eff}(R)$ is the effective pressure 
at the vacuum stellar boundary or at the surface. When $a$ is very small, 
which can eventually be neglected, then this condition reduces to the 
standard case of Buchdahl's limit in GR i.e., $C\lesssim 4/9 =0.444$. For the 
existence of the minimum possible mass of a compact object in MTGs this 
condition demands that the parameter $a$ or $p_{eff}(R)$ must be negative, 
i.e.\ $a<0$ or $p_{eff}(R)<0$ \cite{burikham}. Consequently, this implies 
that the upper bound on compactness in MTGs should be slightly higher than 
the Buchdahl's limit if the parameter $a$ or $p_{eff}(R)<0$ gives the 
non-vanishing value \cite{burikham}.

Thus, the compact stellar objects having compactness larger than $1/3$ and 
smaller than (but very close) the condition \eqref{buchdahl} can emit GW 
echoes at a frequency of tens of hertz. The echo frequencies emitted by such 
highly compact objects can be estimated by using the expression for the 
characteristic echo time. This is the time taken by a massless test particle 
to travel from the unstable light ring to the centre of the star and is 
expressed as \cite{urbano}
\begin{equation}
	\label{eq19}
	\tau_{echo}\equiv\int_0^{3M}\!\!\!\! e^{\,(\lambda(r)-\chi(r))/2}
	\; \mathrm{d}r.
	\end{equation} 
Using the relations for the $\lambda(r)$ and $\chi(r)$ obtained from equation 
\eqref{d}, the echo frequencies can be calculated as $\omega_{echo} \approx \pi/\tau_{echo}$.

%%%%%%%%%%%%%%%%%%%%%%%%%%%%%%%%%%%%%%%%%%%%%%%%%%%%%%%%%%%%%%%%%%%%%%%%%%%%%%%%%%

\section{Numerical results}\label{numerical}
In this section we discuss the numerical results of the work including the 
associated different stability criterion of the resulting stellar solutions. 
As mentioned earlier, we have solved the modified TOV equations \eqref{q} and 
\eqref{r} for the $f(\mathcal{R})$ gravity models together with EoSs and 
supplemented by some boundary conditions. To do so, we have employed 
the fourth order Range-Kutta method. To comment on the viability and stability of
the obtained results we first checked the mass of stellar structures as a function 
of their radius. These  M-R relationships lead ones to confirm whether the 
obtained solutions are feasible or not. Besides the M-R profiles, the 
energy density, pressure, surface redshift, relativistic adiabatic index, 
pulsation modes of stars are also handy tools to comments on the stability of 
stellar solutions. In the following we have discussed all these properties of the 
obtained results of strange star structures. Also we have compared the masses and 
radii of strange stars determined for our cases with some observed candidates of 
strange stars. These observational results are listed in the  Table\;\ref{tab:table1}. 
This table shows the results for 25 possible strange star candidates. The values for 
these masses and radii indicated in this table are taken from different articles 
(detailed references are included for each of the star in the table). 
In Ref.\;\cite{aziz}, the authors provided a list of possible strange star candidates 
along with the experimental results for their observed mass, radius and some other 
physical parameters. In this article we have extended the list by including more 
candidates as listed in Table\;\ref{tab:table1}. We have used these strange star 
candidates to constrain the considered $f(\mathcal{R})$ gravity models in this context. 

\begin{table*}[!h] 
\caption{\label{tab:table1} List of possible strange star candidates.}
\begin{ruledtabular}
\begin{tabular}{cccccccc}
 Star & Mass M & Radius R & References\\[-2pt] 
 &     (in $\mbox{M}_{\odot}$) & (in $\mbox{km}$) \\ \hline
PSR J0740$+$6620	   & $2.072^{+0.067}_{-0.066}$ & $12.39^{+1.30}_{-0.98}$ & \cite{{cromartie}}\\
4U 1636$-$536       & 2.02 $\pm$ 0.12 & 9.6 $\pm$ 0.6     & \cite{kaaret} \\
PSR J0348$+$0432    & 2.01 $\pm$ 0.04 & 12.605 $\pm$ 0.35 &  \cite{antoni}\\
PSR J1614$-$2230    & 1.97 $\pm$ 0.04 & 9.69 $\pm$ 0.2    &  \cite{demo, gango}\\ 
KS 1731$-$260       & 1.8             & 12                &  \cite{abu} \\
Vela X$-$1          & 1.77 $\pm$ 0.08 & 9.56 $\pm$ 0.08   &  \cite{gango, rawls}\\ 
4U 1608$-$52        & 1.74 $\pm$ 0.14 & 9.3 $\pm$ 1.0     &  \cite{poutanen}\\ 
RX J1856.5$-$3754   & 1.7 $\pm$ 0.4   & 11.4 $\pm$ 2      &  \cite{darke} \\ 
4U 1822$-$37        & 1.69 $\pm$ 0.13 & 10                &  \cite{laria}\\
PSR J1903+0327      & 1.667 $\pm$ 0.021&9.438 $\pm$ 0.03  &  \cite{gango, frieri}\\ 
EXO 0748$-$676      & 1.64 $\pm$ 0.38  & $13.8^{+0.6}_{-2.0}$ &  \cite{degenaar}\\
SAX J1808.4$-$3658  & 1.6              & 11               & \cite{li, poutanen}\\
4U 1820$-$30		    & 1.58 $\pm$ 0.06  & 9.1 $\pm$ 0.4& \cite{guver}\\
Cen X$-$3			& 1.49 $\pm$ 0.08  & 9.178 $\pm$ 0.13& \cite{gango, rawls}\\
Cyg X$-$2			& 1.44 $\pm$ 0.06  & 9.0 $\pm$ 0.5& \cite{tita}\\
PSR 1937+21		    & 1.4		       & 6.6& \cite{ren}\\
EXO 1745$-$248		& 1.4		       & 11& \cite{ozel}\\
SAX J1748.9$-$2021	& 1.33 $\pm$ 0.33   & 10.93 $\pm$ 2.09& \cite{tguver}\\
EXO1785$-$248		& 1.3 $\pm$ 0.2	    & 8.849 $\pm$ 0.4& \cite{gango, ozel}\\
LMC X$-$4			& 1.29 $\pm$ 0.05	& 8.831 $\pm$ 0.09& \cite{rawls}\\
4U 1728$-$34			& 1.1		        & 9& \cite{lii}\\
SMC X$-$1			& 1.04 $\pm$ 0.09	& 8.301 $\pm$ 0.2& \cite{aziz}\\
4U 1538$-$52		    & 0.87 $\pm$ 0.07	& 7.866 $\pm$ 0.21& \cite{baker}\\
Her X$-$1			& 0.85 $\pm$ 0.15	& 8.1 $\pm$ 0.41& \cite{gango, abu}\\
PSR B0943+10			& 0.02				& 2.6			& \cite{yue}
\end{tabular}
\end{ruledtabular}
\end{table*}

%%%%%%%%%%%%%%%%%%%%%%%%%%%%%%%%%%%%%%%%%%%%%%%%%%%%%%%%%%%%%%%%%%%%%%%%%%%%%%%%%%%%%%% 

\subsection{Mass-radius profiles and GW echoes}
\begin{figure*}[!h]
        \centerline{
        \includegraphics[scale = 0.27]{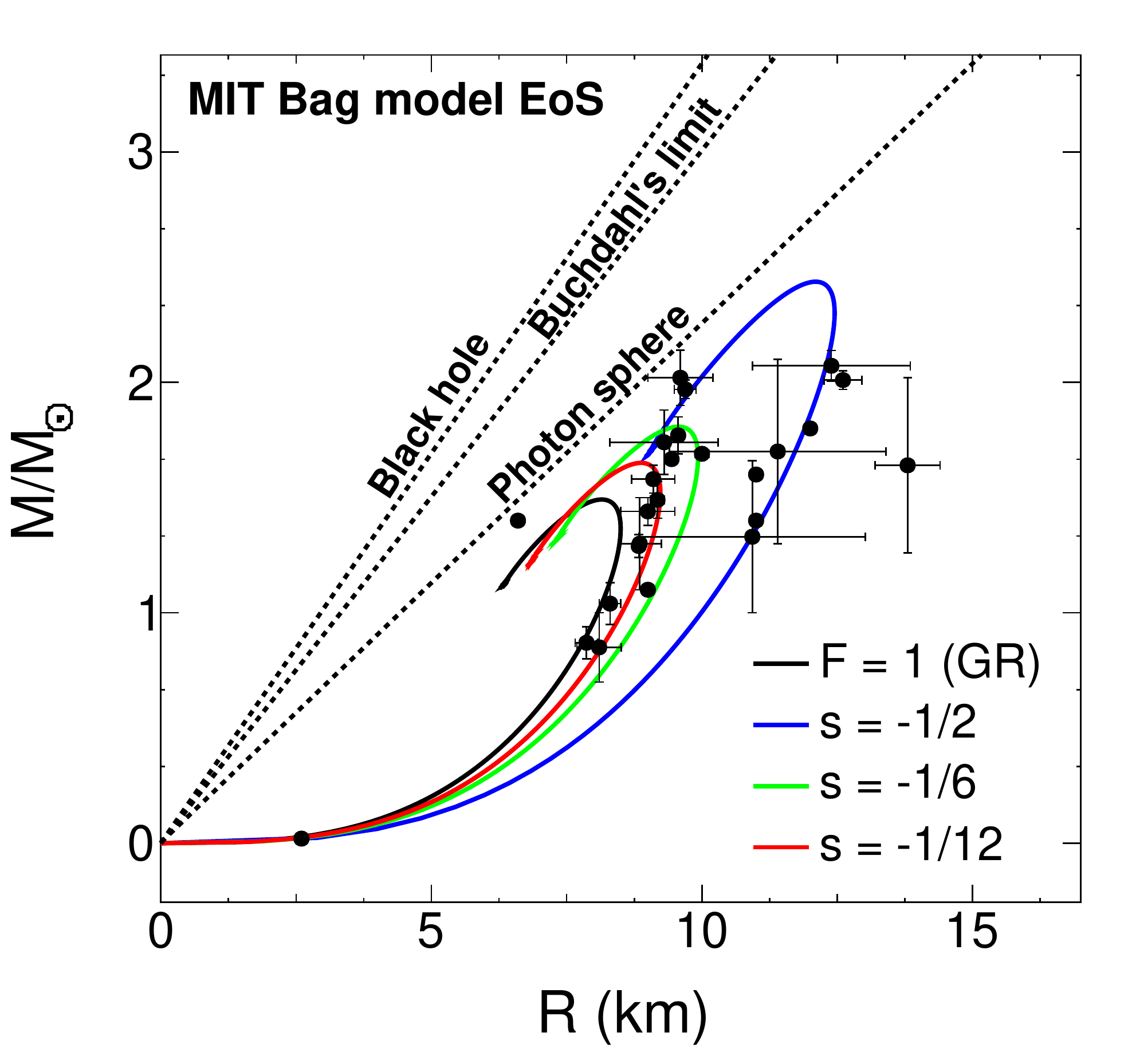}\hspace{0.2cm}
        \includegraphics[scale = 0.27]{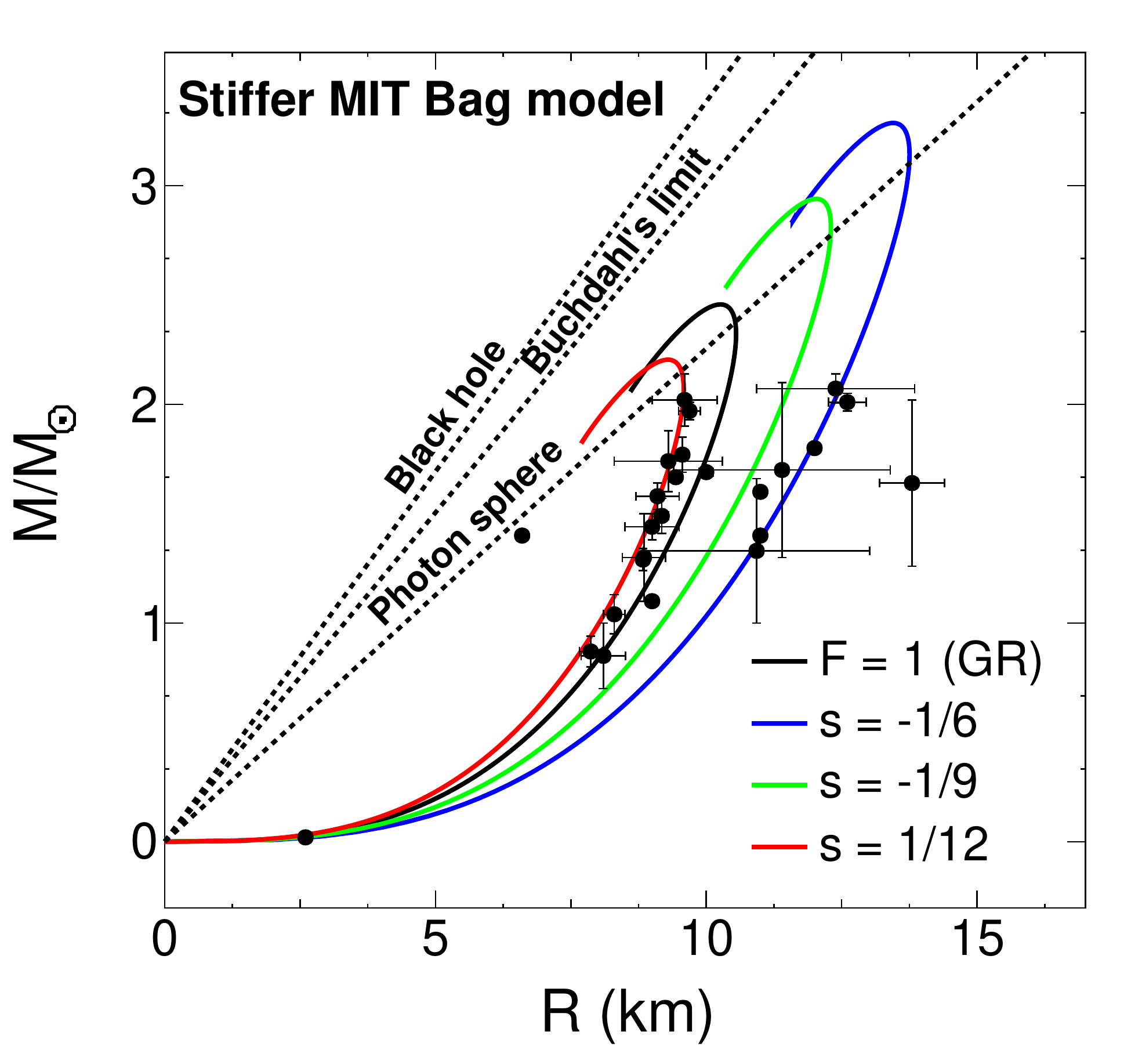}\hspace{0.2cm}
        \includegraphics[scale = 0.27]{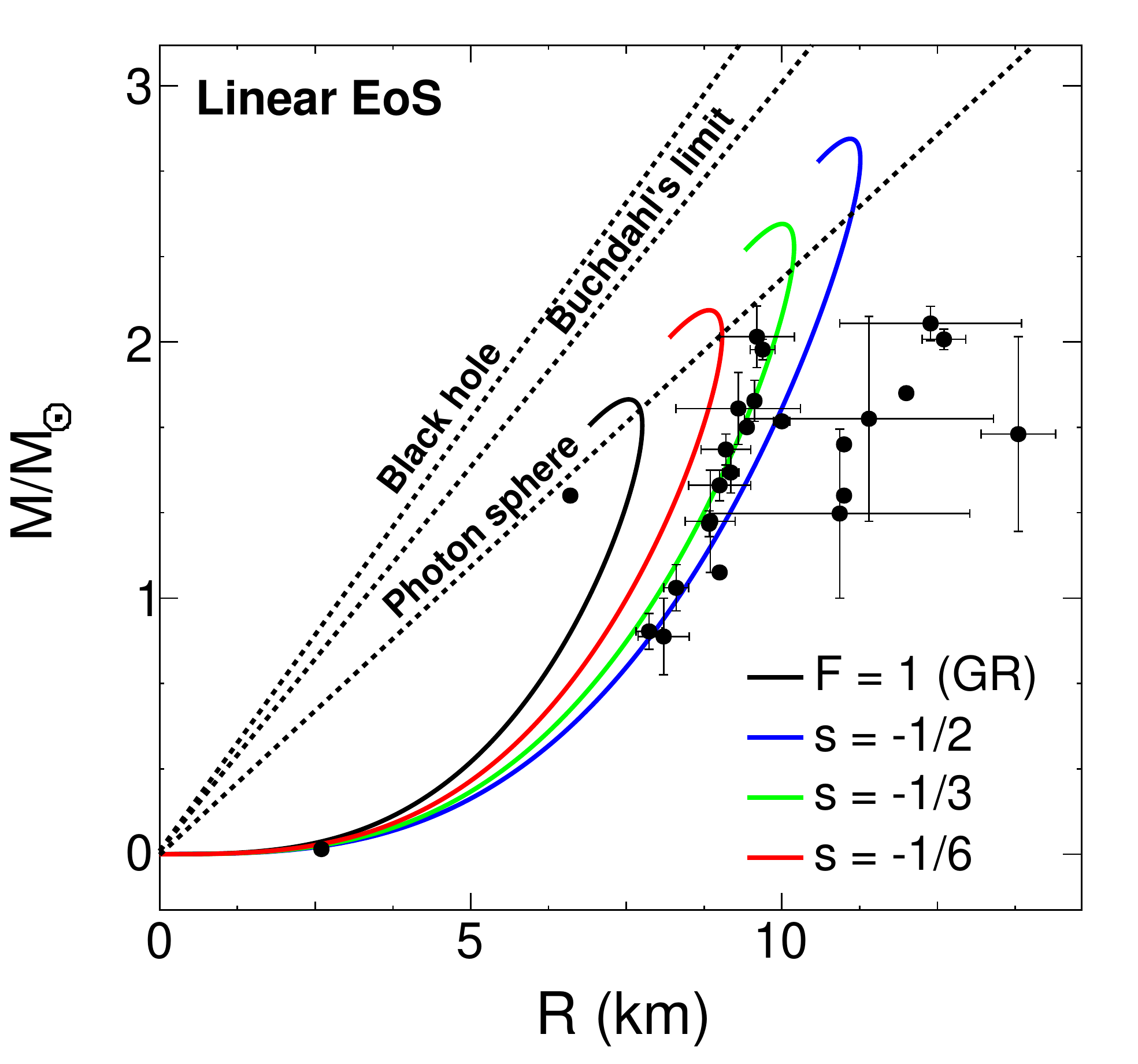}}
        \vspace{-0.3cm}
        \caption{The variation of mass function against the radius of strange
stars in the Starobinsky model for the MIT Bag model EoS (left panel), the 
stiffer MIT Bag model EoS (middle panel) and the linear EoS (right panel) 
respectively with different values of the model parameter $s$. The 
observational data from Table\;\ref{tab:table1} are incorporated together in 
these plots as shown by black dots. Masses of stars are in the unit of Solar 
mass $\mbox{M}_\odot$ and the mass scale of the model $\mathcal{M}_{s}$
is taken as unity. The Buchdahl's limit shown in this and the following
two figures is the maximum compactness limit of a maximum possible massive 
compact stellar object in GR. This limit in MTGs should be of slightly higher
value \cite{burikham}.}
        \label{fig1}
        \end{figure*}

\begin{figure*}[!h]
        \centerline{
        \includegraphics[scale = 0.27]{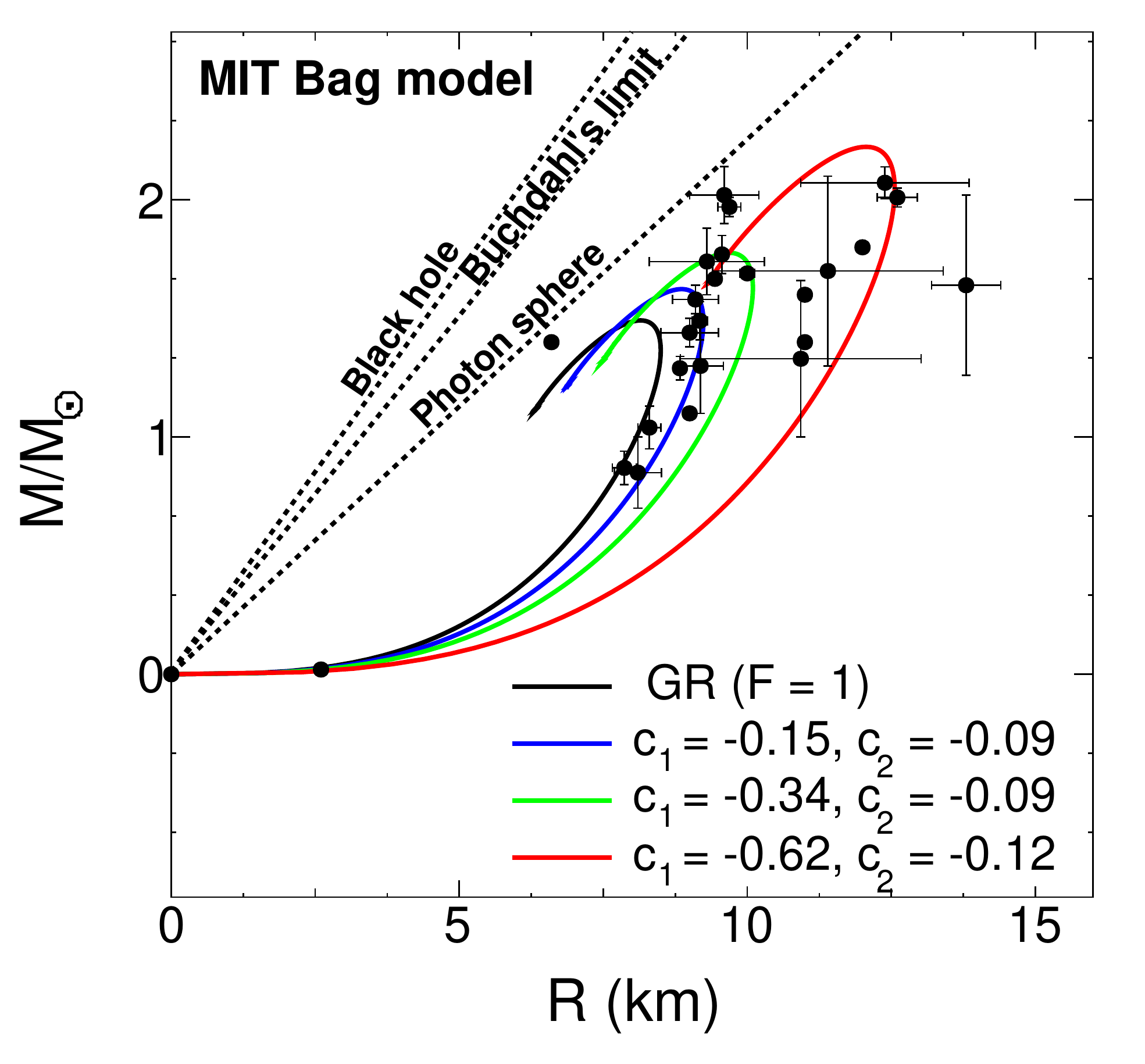}\hspace{0.2cm}
        \includegraphics[scale = 0.27]{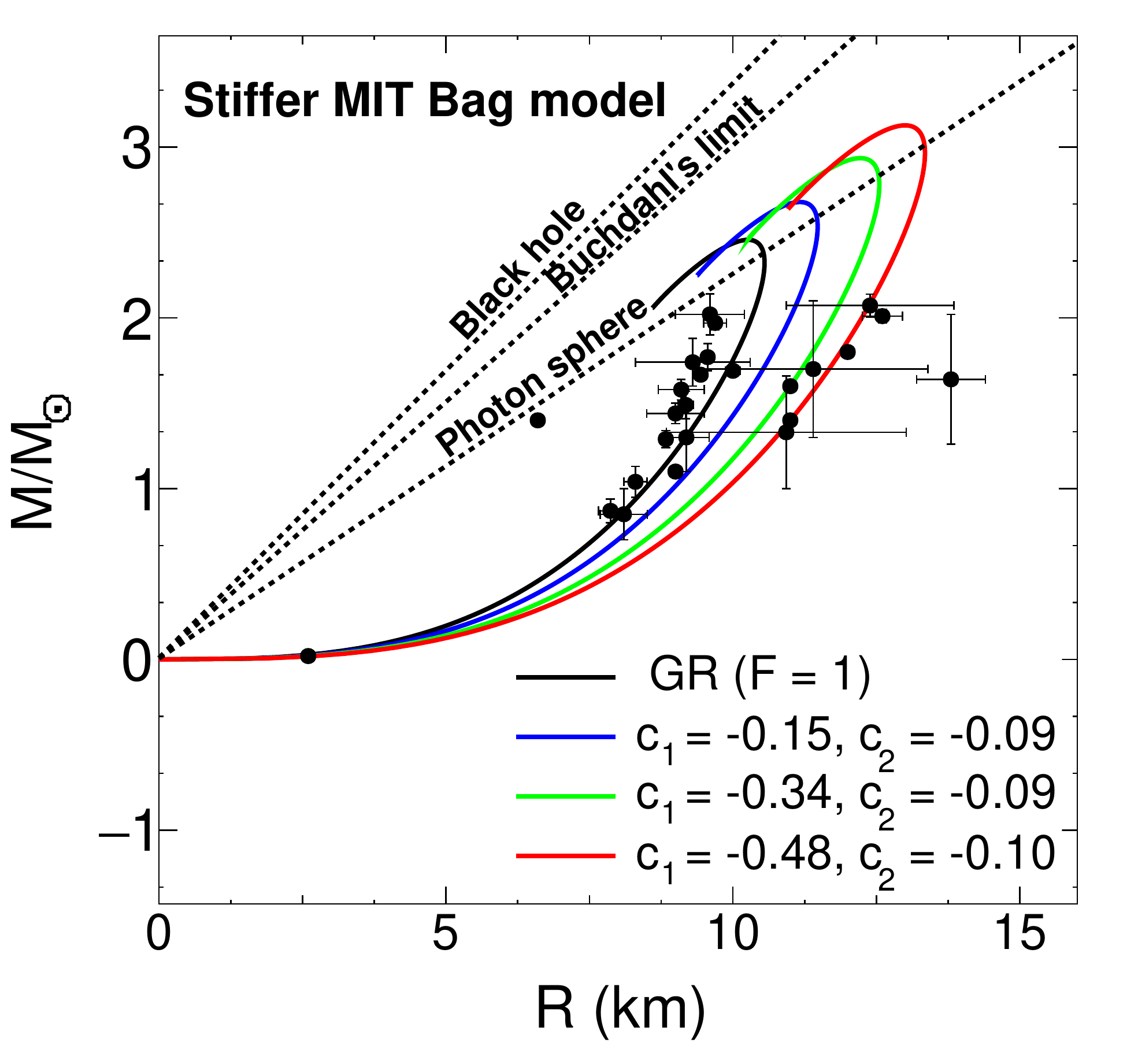}\hspace{0.2cm}
        \includegraphics[scale = 0.27]{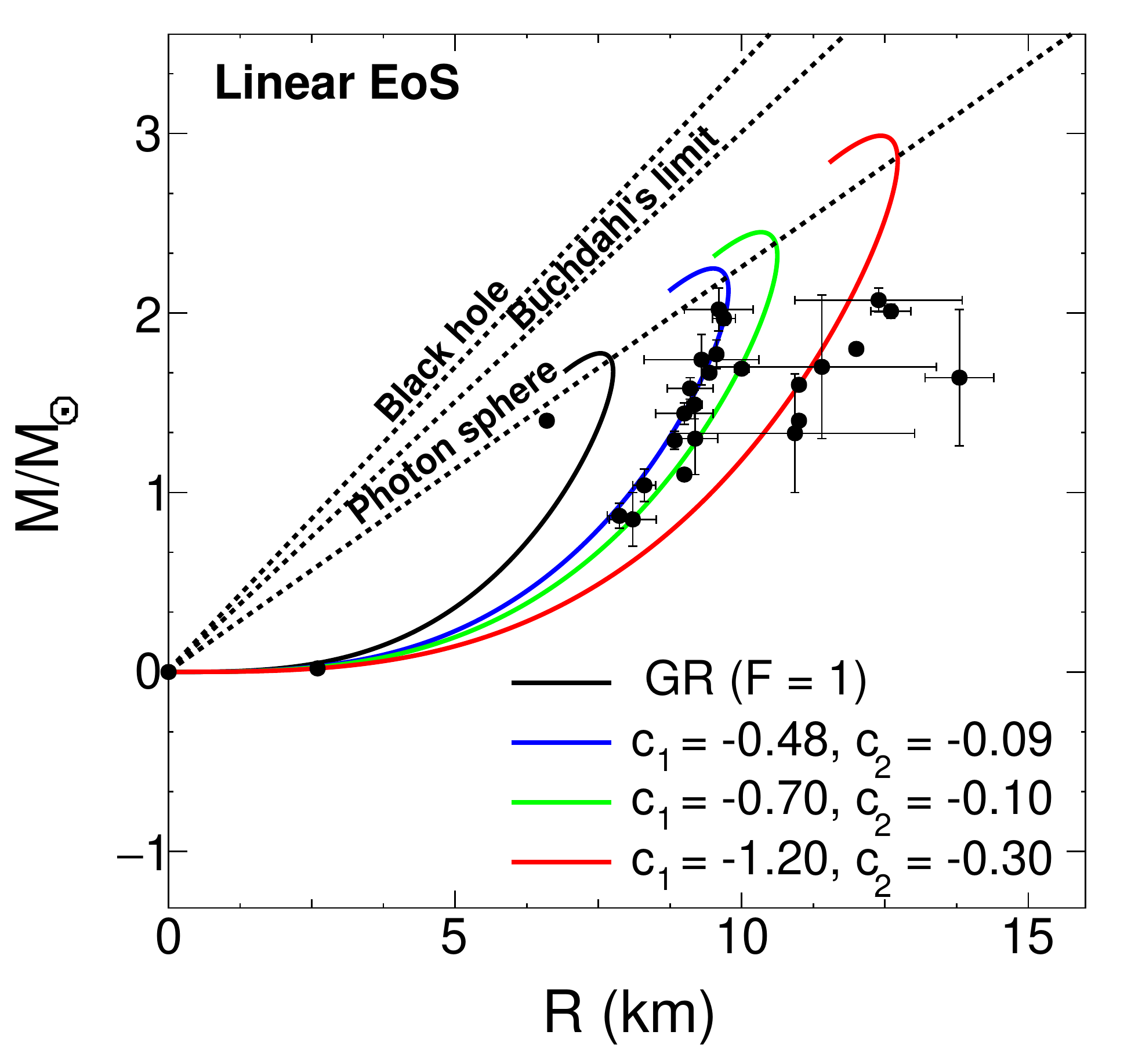}}
        \vspace{-0.3cm}
        \caption{The variation of mass function with respect to radius of
        strange stars in the Hu-Sawicki model for the MIT Bag model EoS 
        (left panel), the stiffer MIT Bag model EoS (middle panel) and the 
        linear EoS (right panel) respectively for different values of the 
        model parameters $c_1$ and $c_2$. The observational data from 
        Table\;\ref{tab:table1} are incorporated together in these plots as 
        shown in black dots. Masses of stars are in the units of solar mass 
        $M_\odot$ and $\mathcal{M}_h$ is taken as unity.}
        \label{fig2}
        \end{figure*}

\begin{figure*}[!h]
        \centerline{
        \includegraphics[scale = 0.27]{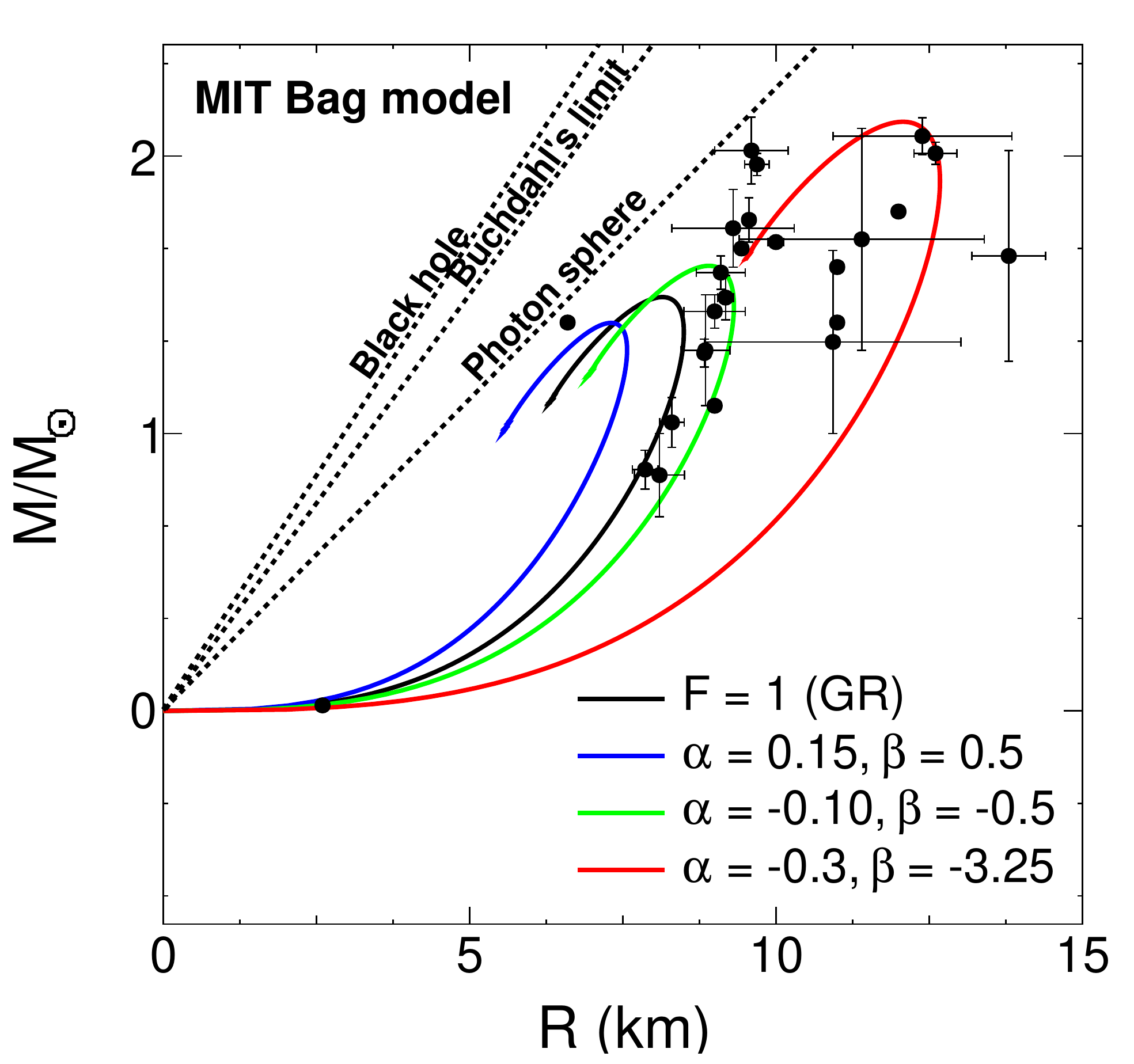}\hspace{0.2cm}
        \includegraphics[scale = 0.27]{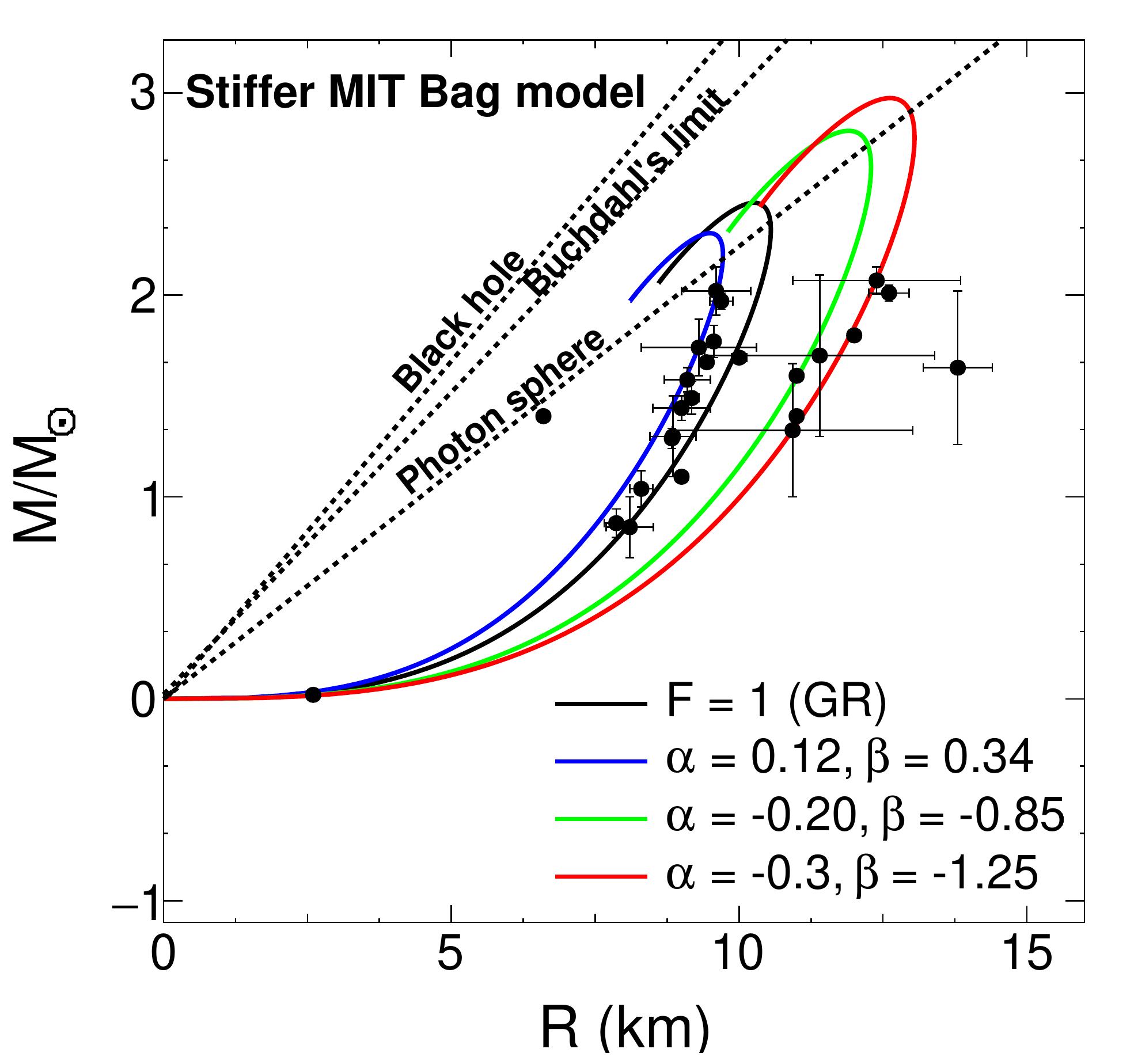}\hspace{0.2cm}
        \includegraphics[scale = 0.27]{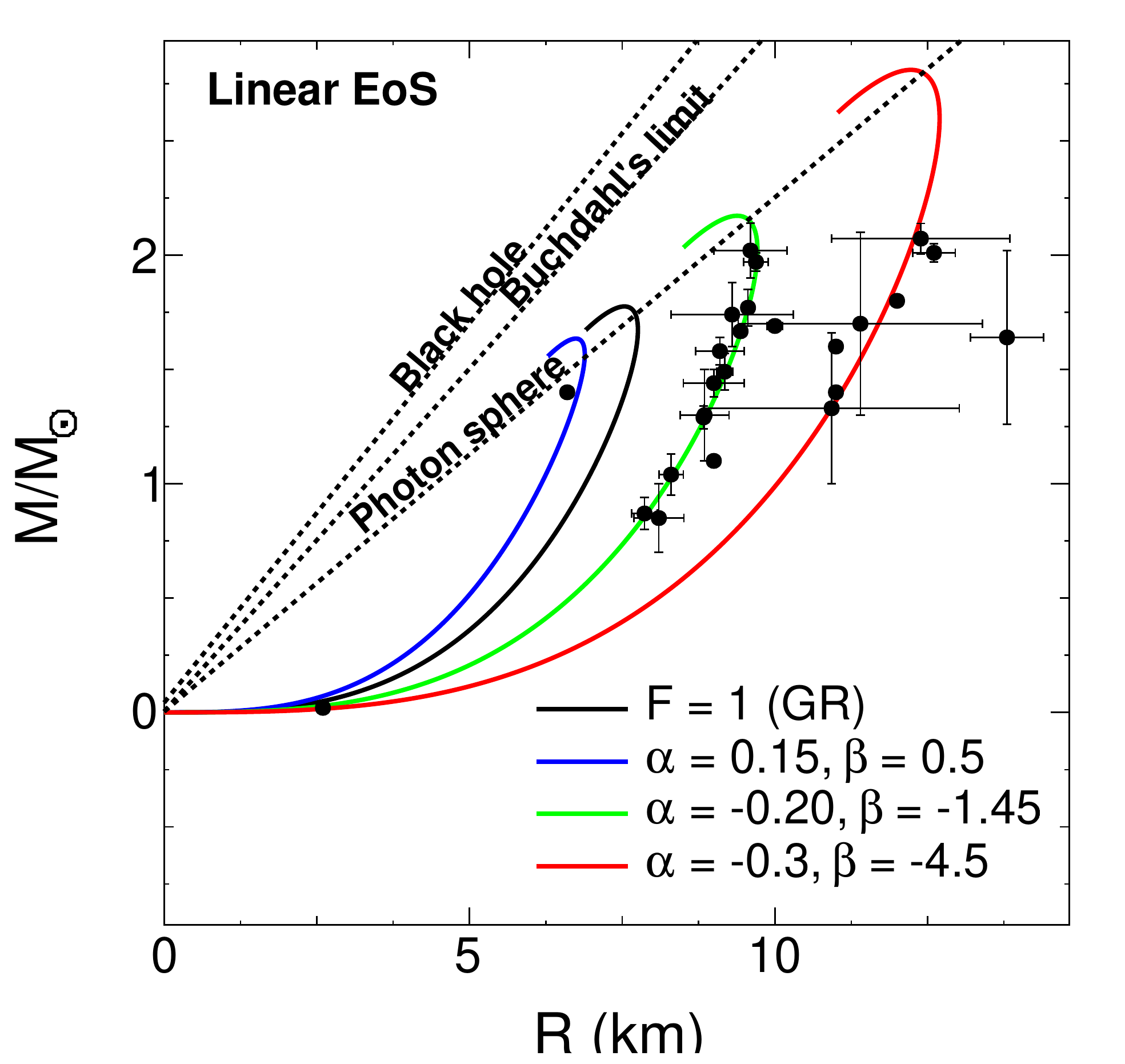}}
        \vspace{-0.3cm}
        \caption{The variation of mass function against the radius of strange 
stars in the Gogoi-Goswami model for the MIT Bag model EoS (left panel), the
stiffer MIT Bag model EoS (middle panel) and the linear EoS (right panel) 
respectively for different values of the model parameter $\alpha$ and $\beta$. 
The observational results from Table\,\ref{tab:table1} are incorporated 
together in these plots as shown by black dots. Masses of stars are in the 
units of solar mass $M_\odot$ and the characteristic curvature 
$\mathcal{R}_c$ of the model is taken as unity.}
        \label{fig3}
        \end{figure*}

The structural behaviours of strange stars are unique in comparison to other 
compact stars like neutron stars and white dwarfs. Figs.\;\ref{fig1}-\ref{fig3} illustrate the mass as a function of the radius $R$ (in km) of strange stars 
in the considered $f(\mathcal{R})$ gravity models for the three standard EoSs. 
In these figures the M-R profiles for the case of GR are also shown in order to 
compare the results altogether. Also these figures are incorporated with the 
standard results of the observational data of compact star candidates listed 
in Table\;\ref{tab:table1}. These data are depicted by the black dots with 
error bar in the figures. These observational data allowed us to comment on 
the viability and the model parameter range of each of the model considered 
here. In Fig.\;\ref{fig1} the M-R curves for strange stars in the 
Starobinsky model are shown. The first panel of this figure corresponds to the 
most general MIT Bag model EoS. From this figure we observed that the strange 
stars predicted by $f(\mathcal{R})=\mathcal{R}-s\mathcal{R}^2/\mathcal{M}_{s}^2$ 
model are mostly larger in mass and radius than that of the GR case when $s$ 
is negative. For more negative values of the model parameter $s$, more massive 
and larger radii configurations are appeared.  However, with the available 
observational data it is thus clear that the values of the model parameter $s$ 
can range from $-1/12$ to $-1/2$, where $s=-1/2$ gives the largest 
configurations corresponding to this EoS. The observed values of the strange 
star candidates are matching for these $s$ values. 
In the second panel of Fig.\ \ref{fig1}, stars with the stiffer MIT Bag model 
EoS are shown. In this case it is observed that unlike the case of MIT Bag 
model EoS, the stiffer form is able to give the stellar structures which are 
capable to cross the photon sphere limit. These structures are now 
contributing to the emission of GW echoes. The GW echo frequency obtained for 
the most stable configuration with $s=-1/6$ is $59.882$ kHz. 
For $s=-1/9$ and $s=1/12$ the estimated echo frequencies are 
$63.765$ kHz and $81.829$ kHz respectively. With the positive
$s$ value we have obtained larger value of echo frequency. For this case we 
found that, the GR case is less matching with the observational data of 
strange star candidates than the stellar structures obtained with the 
Starobinsky model. For this $f(\mathcal{R})$ model with the stiffer EoS, the 
most reasonable good agreement of observational data are seen for the case of
$s=1/12$. Whereas the negative values of $s$  are less likely to depict the 
standard results for this case. Thus it can be inferred that in the case of 
stiffer EoS, a positive $s$ is helpful in describing the observed candidates 
than the negative $s$ values. Again for the stellar configurations in the 
Starobinsky model with the linear EoS, we observed that the stellar solutions
for $s=-1/3$ is found to match pretty well with the observed strange star 
candidates. However, $s$ values greater than $|-1/2|$ and less than $1$ are 
not desirable as these values will lead to more massive compact objects
($M>3M_\odot$). Moreover, in this case the solutions for GR are not matching 
with the observed strange star candidates. Also, stars obtained from this EoS 
are compact enough to emit GW echo frequencies. For stars with $s=-1/2$, the 
characteristic echo frequency corresponding to the maximum mass configuration 
is found to be $55.097$ kHz. The other values of echo frequencies can 
be found in the Table\;\ref{tab:table2}.    
\begin{table*}[!h]
\caption{\label{tab:table2} Physical parameters of strange stars in the 
Starobinsky model for different values of $s$ with $\mathcal{M}_{s} = 1$.}
\begin{ruledtabular}
\begin{tabular}{ccccccccc}
EoSs & $s$ & Radius $\mbox{R}$ & Mass $\mbox{M}$ & Compactness & Redshift & Echo & GWE$^\ast$ & $f$-mode\\[-2pt] 
&  & (in $\mbox{km}$) & (in $\mbox{M}_{\odot}$) & ($\mbox{M/R}$) & $\mbox{Z}$ & time ($\mbox{ms}$) & frequency (kHz)& frequency (kHz)\\ \hline
\multirow{3}{*}{MIT Bag model EoS} 
& -1/2  &  12.101 & 2.437 & 0.298 & 0.600 & -  & - & 7.580\\
& -1/6  &   9.565 & 1.808 & 0.279 & 0.522 & -  & - & 11.318\\
& -1/12 &   8.875 & 1.651 & 0.275 & 0.503 & -  & - & 12.656\\ \hline
\multirow{3}{*}{Stiffer MIT Bag model EoS}
& -1/6  & 12.032 & 2.940 & 0.361 & 0.840 & 0.052 & 59.882 & 5.558\\
& -1/9  & 11.462 & 2.781 & 0.359 & 0.815 & 0.049 & 63.765 & 6.107\\ 
& 1/12  &  9.293 & 2.204 & 0.351 & 0.752 & 0.038 & 81.829 & 8.646\\ \hline
\multirow{3}{*}{Linear EoS}
& -1/2  & 11.097  & 2.793 & 0.372 & 0.679 & 0.057 & 55.097 & 6.206\\
& -1/3  & 10.008  & 2.461 & 0.364 & 0.599 & 0.047 & 66.691 & 7.818 \\ 
& -1/6  &  8.831  & 2.124 & 0.356 & 0.544 & 0.039 & 81.158 & 9.897\\
\end{tabular}
\end{ruledtabular}
$^\ast${\small Gravitational wave echoes}
\end{table*}
In this table the maximum mass and the corresponding radius, compactness, 
surface redshift, characteristic echo time, echo frequency and frequency 
of $f$-mode of oscillation for strange 
stars in the Starobinsky model with three different values of the model 
parameter $s$ within the range allowed by observational data are listed. 
While calculating these values shown in the table (Table\;\ref{tab:table2}) 
and also in the Fig.\;\ref{fig1} the mass scale of Starobinsky model is taken 
as $\mathcal{M}_{s}=1$, in the unit of $\mathcal{R}$. Moreover, choosing a 
particular value of the model parameter $s=-1/6$ and with different values of 
$\mathcal{M}_{s}$, all the said physical parameters related with the obtained 
strange star configurations are also calculated. These results are summarized 
in Table\;\ref{tab:table3}. It is observed that with the increasing values of 
$\mathcal{M}_{s}$, the mass, radius and compactness decreases, whereas the 
echo frequency increases with increase in $\mathcal{M}_{s}$ values, for all 
the considered cases.   
\begin{table*}[!h]
\caption{\label{tab:table3} Physical parameters of strange stars in the 
Starobinsky model for different $\mathcal{M}_s$ values with $s=-1/6$.}
\begin{ruledtabular}
\begin{tabular}{ccccccc}
EoSs & $\mathcal{M}^2_s$ & Radius $\mbox{R}$ & Mass $\mbox{M}$ & Compactness & Echo & GWE\\[-2pt] 
     &(in units of $\mathcal{R}$) & (in $\mbox{km}$) & (in $\mbox{M}_\odot$) & ($\mbox{M/R}$) & time ($\mbox{ms}$) & frequency (kHz)\\ \hline
\multirow{3}{*}{MIT Bag model EoS} 
& 2 & 8.875 & 1.651 & 0.275 & -  & - \\
& 1 & 9.565 & 1.808 & 0.279 & -  & - \\
& 2/3 & 10.229 & 1.964 & 0.284   & -  & - \\ \hline
\multirow{3}{*}{Stiffer MIT Bag model EoS}
& 2  & 11.171 & 2.701 & 0.358 & 0.047 & 65.866 \\ 
& 1  & 12.032 & 2.940 & 0.361 & 0.052 & 59.882 \\
&2/3 & 12.854 & 3.175 & 0.365 & 0.057 & 54.690\\ \hline
\multirow{3}{*}{Linear EoS}
& 2  &  8.201 & 1.952 & 0.352 & 0.035 &  90.119 \\
& 1  &  8.831 & 2.124 & 0.356 & 0.039 &  81.158 \\
& 2/3& 9.432  & 2.293 & 0.360 & 0.042 &  73.459 \\ 
\end{tabular}
\end{ruledtabular}
\end{table*}

For the Hu-Sawicki model of $f(\mathcal{R})$ gravity the M-R profile are shown 
in Fig.\,\ref{fig2}. While solving this system for the Hu-Sawicki model, the 
mass scale value for this model is chosen to be $\mathcal{M}_h=1$ in the units 
of $\mathcal{R}$ for this particular model. The first, second and third panel 
of this figure correspond to the MIT Bag model EoS, stiffer MIT Bag model EoS 
and the linear EoS respectively. For the case of MIT Bag model EoS with three 
sets of model parameters $c_1$ and $c_2$, it is seen that the resulting 
configurations lies below the photon sphere limit and with a compactness of 
$\approx 0.27$. Thus these configurations cannot produce GW echoes. 
Incorporating Table\,\ref{tab:table1} for the observed strange star candidates 
in these M-R profile clarifies that for the considered $c_1$ and $c_2$ values 
the resulting compact stars are within the observable range. As shown in the 
second panel of this figure, the M-R profiles given by the stiffer MIT Bag 
model EoS are compact enough to produce echoes. However for this typical case, 
the observed results are not found to match with the stars predicted by the GR 
case. For more negative values of $c_1$ and $c_2$, stars with larger masses 
and radii are found. Similar to the case of stiffer MIT Bag model EoS, the 
stars corresponding to linear EoSs are also ultra-compact in nature. Thus they 
can echo GWs falling on their stellar surfaces. In this case the observed 
strange star candidates are found to match with the stars given by the model 
parameter values $c_1 = -0.48$, $c_2 = -0.09$. For more smaller values of 
these two parameters larger configurations are obtained. A more details about 
the mass, radius, compactness and other physical parameters are listed in 
Table\,\ref{tab:table4}. From this table it is inferred that unlike the MIT 
Bag model (both softer and stiffer forms) the compactness of stellar structures
 increase gradually for the case of linear EoS, in the considered range of 
model parameter values. Also decreasing $c_1$ and $c_2$ values implies 
decrease in the echo frequencies of these stars. Varying the mass scale 
$\mathcal{M}_h$ also shows a mass scale dependent nature of stars. This can 
be seen from Table\,\ref{tab:table5}. From this table it is clear that for 
smaller $\mathcal{M}_h$ values the mass and radius of the most stable 
configuration are increasing slightly, while maintaining the constant 
compactness. However for the case of stiffer MIT Bag model and linear EoS, the 
echo frequencies decrease gradually with decrease in the $\mathcal{M}_h$ values.
\begin{table*}[!h]
\caption{\label{tab:table4} Physical parameters of strange stars predicted by 
the Hu-Sawicki model for different values of the model parameters $c_1$ and 
$c_2$, but with $\mathcal{M}_h = 1$.}
\begin{ruledtabular}
\begin{tabular}{cccccccccc}
EoSs & $c_{1}$ & $c_{2}$ & Radius $\mbox{R}$ & Mass $\mbox{M}$ & Compactness & Redshift & Echo & GWE & $f$-mode\\[-2pt] 
& & & (in $\mbox{km}$) & (in $\mbox{M}_{\odot}$) & ($\mbox{M/R}$) & $\mbox{Z}$ & time ($\mbox{ms}$) & frequency (kHz)& frequency (kHz)\\ \hline
\multirow{3}{*}{MIT Bag model EoS} 
& -0.15 & -0.09 & 8.859 & 1.624 & 0.271 & 0.483 & - & - & 3.674\\
& -0.34 & -0.09 & 9.687 & 1.771 & 0.271 & 0.484 & - & - & 3.043\\
& -0.62 & -0.12 & 10.957 & 2.014 & 0.271 & 0.491 & - & - & 2.169\\ \hline
\multirow{3}{*}{Stiffer MIT Bag model EoS}
& -0.15 & -0.09 & 11.168 & 2.678 & 0.355 & 0.783 & 0.040 & 66.880 & 6.398\\
& -0.34 & -0.09 & 12.220 & 2.935 & 0.355 & 0.794 & 0.052 & 60.953 & 5.373\\ 
& -0.48 & -0.10 & 12.999 & 3.127 & 0.355 & 0.799 & 0.055 & 57.146 & 4.695\\ \hline
\multirow{3}{*}{Linear EoS}
& -0.48 & -0.09 &  9.502  & 2.249 & 0.350 & 0.498 & 0.037 & 79.069 & 8.637\\
& -0.70 & -0.10 & 10.337  & 2.450 & 0.351 & 0.499 & 0.043 & 72.298 & 7.556\\ 
& -1.20 & -0.30 & 12.426  & 2.988 & 0.355 & 0.546 & 0.054 & 57.705 & 5.268\\
\end{tabular}
\end{ruledtabular}
\end{table*}
\begin{table*}[!h]
\caption{\label{tab:table5} Physical parameters of strange stars predicted by 
the Hu-Sawicki model for different $\mathcal{M}_h$ values, but with $c_1=-0.15$, $c_2=-0.09$.}
\begin{ruledtabular}
\begin{tabular}{ccccccc}
EoSs & $\mathcal{M}^2_h$ & Radius $\mbox{R}$ & Mass $\mbox{M}$ & Compactness & Echo & GWE\\[-2pt] 
     & (in units of $\mathcal{R}$) &(in $\mbox{km}$) & (in $\mbox{M}_{\odot}$) & ($\mbox{M/R}$) & time ($\mbox{ms}$) & frequency (kHz)\\ \hline
\multirow{3}{*}{MIT Bag model EoS} 
& 1.05 &  8.849 & 1.621 & 0.271 & - & - \\
&    1 &  8.859 & 1.624 & 0.271 &  - & - \\
& 0.95 &  8.871 & 1.626 & 0.271 &  - & - \\ \hline
\multirow{3}{*}{Stiffer MIT Bag model EoS}
& 1.05 & 11.154 & 2.674 & 0.355 & 0.047 & 66.989 \\
&    1 & 11.168 & 2.678 & 0.355 & 0.047 & 66.880\\
& 0.95 & 11.184 & 2.682 & 0.355 & 0.047 & 66.764 \\ \hline
\multirow{3}{*}{Linear EoS}
& 1.05  & 8.190  & 1.932 & 0.349 & 0.034 & 92.423\\
&    1  & 8.201  & 1.935 & 0.349 & 0.034 & 92.255\\ 
& 0.95  & 8.212  & 1.938 & 0.349 & 0.034 & 92.075\\
\end{tabular}
\end{ruledtabular}
\end{table*}  

For the new $f(\mathcal{R})$ gravity model i.e.\ for the Gogoi-Goswami model, 
the M-R profiles are shown in Fig.\,\ref{fig3}. In this model also the MIT Bag 
model EoS generally gives stars with compactness around $0.27$ and hence such 
compact objects cannot produce GW echoes. This is shown in the first panel of 
Fig.\,\ref{fig3}. For all the considered $\alpha$ and $\beta$ values, all the 
obtained configurations are lying below the limiting line i.e., below the 
photon sphere limit. In this case the stars given by the GR and for 
$\alpha=-0.10$, $\beta=-0.5$ are in agreement with some of the observed data 
of the strange star candidates. For $\alpha=-0.30$, $\beta=-3.25$, stars with 
configurations of maximum mass and radius are obtained, and they are laying 
within the observed data of similar type. For this gravity model with the MIT 
Bag model EoS we observed a rapid drop in compactness with smaller $\alpha$ and $\beta$ values. For the case of stiffer EoS (second panel of Fig.\,\ref{fig3}), the resulting configurations are with the compactness $>0.33$. Smaller 
$\alpha$ and $\beta$ values are giving larger configurations (this is 
applicable to all three EoSs). It is found that the combination of parameters 
$\alpha=0.12$ and $\beta=0.34$ represents stars with a more realistic 
characteristics for the stiffer EoS. 
%Similar to this case, for stars with the linear EoS smaller values of 
%$\alpha$ and $\beta$ are resulting stars with larger structures. This 
%variations can be visualize in the third panel of Fig.\,\ref{fig3}. 
For the linear EoS as shown in the third panel of Fig.\,\ref{fig3}, the 
solutions in GR are virtually not matching with the observed strange star 
candidates (in fact this is the situation for this EoS in all three figures). 
However, for the Gogoi-Goswami model the stellar configurations obtained with 
the two model parameter sets $\alpha=-0.20$, $\beta=-1.45$ and 
$\alpha=-0.3$, $\beta=-4.5$ are in good agreement with the observed results. 
These results are summarized in Table\,\ref{tab:table6}. As it can be 
visualize from the second and third panels of Fig.\,\ref{fig3} that the 
resulting structures are crossing the photon sphere, so they can emit GW echo 
frequencies. The calculated echo frequencies are listed in 
Table\,\ref{tab:table6}. It is seen that with the decreasing $\alpha$ and 
$\beta$ values the echo frequencies are getting smaller for both the EoSs. 
Here the solutions of modified TOV equations are obtained by considering the 
characteristic curvature constant $\mathcal{R}_c=1$ when expressed in units 
of $\mathcal{R}$. For different $\mathcal{R}_c$ values the dependence on the 
stellar structure are shown in Table\,\ref{tab:table7}. It is clear from the 
table that when the value of $\mathcal{R}_c$ decreases, the characteristic 
echo frequency also decreases. However, the compactness of these stars are 
found to increase with the decrease in $\mathcal{R}_c$ values for all the 
considered EoSs.
\begin{table*}[!h]
\caption{\label{tab:table6} Physical parameters of strange stars predicted by
the Gogoi-Goswami model for different values of the parameters $\alpha$ and 
$\beta$ and with $\mathcal{R}_c = 1$.}
\begin{ruledtabular}
\begin{tabular}{ccccccccccc}
EoSs & $\alpha$ & $\beta$  & Radius $\mbox{R}$ & Mass $\mbox{M}$ & Compactness & Redshift & Echo & GWE & $f$-mode\\[-2pt] 
& & & (in $\mbox{km}$) & (in $\mbox{M}_{\odot}$) & ($\mbox{M/R}$) & $\mbox{Z}$ & time ($\mbox{ms}$) & frequency (kHz)&  frequency (kHz)\\ \hline
\multirow{3}{*}{MIT Bag model EoS} 
& 0.15 & 0.5  &  7.309 & 1.399  & 0.283 & 0.531 & - & - & 4.983\\
& -0.10 & -0.5 &  8.901 & 1.604  & 0.277 & 0.462 & - & - & 3.673\\
& -0.30 & -3.25 &  12.074 & 2.124  & 0.260 & 0.438 & - & - & 1.490\\ \hline
\multirow{3}{*}{Stiffer MIT Bag model EoS}
&  0.12 &  0.34 &  9.492 & 2.306 & 0.359 & 0.817 & 0.041 & 76.804 & 8.394\\
& -0.20 & -0.85 & 11.904 & 2.813 & 0.349 & 0.752 & 0.049 & 64.235 & 5.621\\ 
& -0.30 & -1.25 & 12.623 & 2.975 & 0.348 & 0.740 & 0.052 & 60.808 & 4.952\\ \hline
\multirow{3}{*}{Linear EoS}
& 0.15 & 0.5  &  6.740 & 1.635 & 0.359 & 0.566 & 0.030 & 103.557 & 13.017\\
& -0.20 & -1.45 &  9.382 & 2.172 & 0.342 & 0.434 & 0.037 & 84.853 & 9.210\\ 
& -0.30 & -4.5 &  12.227 & 2.810 & 0.340 & 0.406 & 0.047 & 66.318 & 6.074\\
\end{tabular}
\end{ruledtabular}
\end{table*}      
\begin{table*}[!h]
\caption{\label{tab:table7} Physical parameters of strange stars predicted by
the Gogoi-Goswami model for different $\mathcal{R}_{c}$ values and with 
$\alpha=0.15$, $\beta=0.5$.}
\begin{ruledtabular}
\begin{tabular}{cccccccc}
EoSs & $\mathcal{R}_{c}$ & Radius $\mbox{R}$ & Mass $\mbox{M}$ & Compactness & Echo & GWE\\[-2pt] 
     &(in units of $\mathcal{R}$) & (in $\mbox{km}$)  & (in $M_{\odot}$) & ($\mbox{M/R}$) & time ($\mbox{ms}$) & frequency (kHz)\\ \hline
\multirow{3}{*}{MIT Bag model EoS} 
& 2   &  6.666 & 1.252  & 0.278 & - & -  \\
& 1   &  7.309 & 1.399  & 0.283 & - & -  \\
& 2/3 &  7.801 & 1.511  & 0.286 & - & -  \\ \hline     
\multirow{3}{*}{Stiffer MIT Bag model EoS}
&  2  &  8.387 & 2.041 & 0.360 & 0.036 & 86.656  \\
&  1  &  9.185 & 2.264 & 0.364 & 0.041 & 76.957  \\ 
& 2/3 &  9.796 & 2.435 & 0.368 & 0.049 & 62.836  \\ \hline
\multirow{3}{*}{Linear EoS}
& 2   & 6.157  & 1.474 & 0.354 & 0.025 & 127.571 \\ 
& 1   & 6.740  & 1.635 & 0.359 & 0.030 & 103.557 \\
& 2/3 & 7.187  & 1.758 & 0.362 & 0.032 & 99.597  \\ 
\end{tabular}
\end{ruledtabular}
\end{table*}

As mentioned earlier the prime objective of the present study is to 
explore the astrophysical implications of the considered three well-behaved 
$f(\mathcal{R})$ models in the Palatini formalism in the perspectives of 
predictions of strange star's behaviours. These models are considered as 
well-behaved models in the sense of their behaviours in the cosmological
perspectives \cite{gg,ijmpd}. Thus for a better picture of the behaviours of
these models in the astrophysical aspects of our study in comparison to their
behaviours in the cosmological domain, we have made a comparative analysis of 
the ranges of parameters obtained from this study for each model with their 
corresponding cosmological ranges obtained from the literature, which is 
shown in Table\,\ref{tab:table8}. For the case of Starobinsky model of the form 
\eqref{u} with $n=1$, the constraint on the model parameter $s$ was reported 
earlier as $s<8/3 \sqrt{3}\approx4.618$ \cite{staro}. Using the cosmic 
expansion and the structure growth data, the constrained value of this model 
parameter is reported as $s^{-1}={1.460}^{+0.47}_{-0.52}$ \cite{bessa}. The 
viable range obtained from our study for this model's parameter with three 
EoSs are shown together with the constrained range reported in 
Ref.\,\cite{staro} in Table~\ref{tab:table8}. From this it is clear that the 
range suitable for strange stars are within the cosmological limit of the 
model's parameter $s$. Cosmological consequences of the Hu-Sawicki model was 
discussed earlier in the 
Ref.\,\cite{santos}. For this gravity model a set of constraints and 
best-fitted values for the model parameters $c_1$ and $c_2$ were reported 
therein. Now the sets of the parameters $c_1$ and $c_2$ considered in this 
study are in the compatible range with that of the values discussed in 
\cite{santos}. For the case of Gogoi-Goswami $f(\mathcal{R})$ gravity model a 
set of constrained parameters $\alpha$ and $\beta$ are discussed in \cite{gg}.
Keeping in mind these cosmologically constrained values we have used different 
sets of the model parameters which are found interesting in this study. 
However rather than using the positive values of this model parameters we have 
obtained that stellar solutions with negative values of $\alpha$ and $\beta$ 
are more interesting from the present perspective of the study.
\begin{table*}[!h]
\caption{\label{tab:table8} Astrophysical and cosmological ranges of the model parameters for MIT Bag model, stiffer MIT Bag model and linear EoS.}
\begin{ruledtabular}
\begin{tabular}{cccccc}
\multirow{2}{*}{$f(\mathcal{R})$ gravity models} & 
\multirow{2}{*}{Model parameters} & 
\multicolumn{3}{c}{Astrophysical range from this study} & 
\multirow{2}{*}{Cosmological range} \\
 &  & MIT Bag model EoS & Stiffer MIT Bag model & Linear EoS &  \\ \hline
Starobinsky model & $s$ & $-0.5$ to $-0.083$ & $-0.166$ to $0.083$ & $-0.5$ to $-0.167$ &  $<4.618$ \cite{staro}\\ \hline
\multirow{2}{*}{Hu-Sawicki model} 
& $c_{1}$ & $-0.62$ to $-0.15$ & $-0.48$ to $-0.15$ & $-1.20$ to $-0.48$ & $-12.64$ to $2.23$ \cite{santos}\\ 
& $c_{2}$ & $-0.12$ to $-0.09$ & $-0.10$ to $-0.09$ & $-0.30$ to $-0.09$ & $-0.85$ to $0.08$ \cite{santos}\\ \hline
\multirow{2}{*}{Gogoi-Goswami model} 
& $\alpha$ & $-0.30$ to $0.15$  & $-0.30$ to $0.12$ & $-0.30$ to $0.15$ & $0.045$ to $0.300$ \cite{gg}\\
& $\beta$ &$-3.25$ to $0.5$  & $-1.25$ to $0.34$  & $-4.5$ to $0.5$  &  $0.500$ to $0.850$          \cite{gg}
\end{tabular}
\end{ruledtabular}
\end{table*}
                
\subsection{Stability analysis}
\subsubsection{$f$-mode of radial perturbations}
In order to study the stability against radial perturbations of the compact 
stars in our configurations we need to calculate the fundamental $f$-mode 
frequencies corresponding to the radial oscillations for all those 
configurations. The dynamical stability against the radial adiabatic 
perturbation of stellar system are studied in a plethora of articles 
\cite{hh,hs,ipser,chan,jb}. The necessary and sufficient condition for 
stability against radial perturbations of any physical system is that the 
eigenfrequency of the lowest normal mode must be real \cite{hh}. In order 
to calculate these radial oscillation frequencies, we follow the 
Ref.\,\cite{pretel2}, where the authors considered the two-fluid formalism. 
Under this formalism it is considered that the star is composed of two fluids, 
the perfect fluid and the curvature fluid. With this %two
idea, we can rewrite the field equation from Eq.\,\eqref{s} as 
\begin{equation}
G_{\mu\nu}=8\pi(T_{\mu\nu}^m+T_{\mu\nu}^c),
\end{equation}
where $T_{\mu\nu}^m$ is the usual matter energy-momentum tensor scaled by a 
factor $F_\rho$ as $\dfrac{T_{\mu\nu}}{F_{\rho}}$ and $T_{\mu\nu}^c$ is the 
curvature fluid with the form: $-\dfrac{\Lambda_{\rho}g_{\mu\nu}}{8\pi}$. This 
consideration gives us the privilege to use the standard equations of radial 
pulsations as discussed in Ref.\,\cite{pretel2}. However, one should note that 
these oscillation equations will result different oscillation modes as the 
configuration is highly dependent on the $f(\mathcal{R})$ gravity models which
is followed from the modified TOV Eqs.\,\eqref{q}-\eqref{r}.

The study of radial pulsation of stellar objects was pioneered by Chandrasekhar 
\cite{Chandrasekhar1964a,Chandrasekhar1964b}. The pulsation equations are 
coupled first order differential equation of Sturm-Liouville type and can be 
written as \cite{jb,vath}
\begin{equation}
	\label{eq11}
	\small {\dfrac{d\xi}{dr} = -\, \dfrac{1}{r}\left(3\xi + \dfrac{\eta}{\Gamma}
	\right) - \dfrac{dp}{dr}\dfrac{\xi}{p+\rho}},
	\end{equation}
	\begin{equation}
	\label{eq12}
	\dfrac{d\eta}{dr} = \xi\left[\omega^{2}r\left(1+\dfrac{\rho}{p}
	\right)e^{\lambda-\chi}-\dfrac{dp}{dr}\dfrac{4}{p}-8\pi(p+
	\rho)\,r\,e^{\lambda}+\;\dfrac{r}{p\,(p+\rho)}{\left(\dfrac{dp}{dr}\right)}^{2}
	\right] +\eta\left[-\,\dfrac{\rho}{p\,(p+\rho)}\dfrac{dp}{dr}-4\pi(p+\rho)\,r
	\,e^{\lambda}\right],
	\end{equation}
where $\xi= \Delta r/r$ and $\eta = \Delta p/p$ are two dimensionless 
variables. $\Delta r$ is the radial perturbation and $\Delta p$ is the 
corresponding Lagrangian perturbations of the pressure. $\Gamma$ is the 
relativistic adiabatic index. It can be defined as \cite{chan} 
\begin{equation}
\label{hh}
\Gamma=\dfrac{p+\rho}{p}\,\dfrac{dp}{d\rho}.
\end{equation} and $\omega$ is 
the eigenfrequency of vibration. This coupled equations are supported by two 
boundary conditions: at $r=0$, $\eta=-3\,\gamma\,\xi$ and $\Delta p=0$ at 
$r\rightarrow R_{eff}$, $R_{eff}$ is the radial distance where the 
total pressure goes to zero. This pulsation equations has real eigenvalues 
$\omega_{0}^{2}<\omega_{1}^{2}< \hdots <\omega_{n}^{2}< \hdots$, and 
$\omega_{n}$ are the eigenfrequencies of oscillations for $n$ nodes. The mode 
corresponding to $n=0$ is called the fundamental or $f$-mode. When 
$\omega^{2}<0$, the frequency $\omega$ becomes imaginary and it corresponds to an 
unstable state.

By solving these pulsation equations for all the considered cases of this study, the 
lowest order mode of radial oscillation i.e., $f$-mode oscillation frequencies are 
computed and are listed in Tables: \ref{tab:table2},  \ref{tab:table4}, 
\ref{tab:table6} for Starobinsky model, Hu-Sawicki model and Gogoi-Goswami model 
respectively. All these frequencies are in the kHz range and are found to have real 
values and eventually signifies the stability of all the considered configurations.
        
\subsubsection{Relativistic adiabatic index}
An important and necessary quantity to describe and check a region of stability 
of relativistic isotropic fluid sphere is the relativistic adiabatic index 
$\Gamma$. For the spherically symmetric spacetime with a perfect fluid 
Chandrasekhar \cite{Chandrasekhar1964a} explored first the instability regimes 
using this adiabatic index. He gave a particular numerical value of the 
adiabatic index i.e. $\Gamma = 4/3$, whose violation with higher $\Gamma$ 
values indicates a stable star configuration. In 1975 Heintzmann and 
Hillebrandt \cite{hh} also proposed that isotropic compact star models are 
stable for $\Gamma > 4/3$ throughout the stellar interior. Basically, 
this stability criteria was reported for the case of compact Newtonian stars 
\cite{bondi}. 
Though the relativistic effects for the theory of stellar oscillations 
are negligible, yet for highly compact objects like neutron stars and strange 
stars the relativistic limits should be taken into consideration rather than 
the Newtonian limit. For relativistic cases the system becomes more 
unstable due to the regenerative effect of pressure \cite{sarkar,erre}. 
However, recently for a stable and relativistic dynamical system, the same 
condition $\Gamma > 4/3$ has been considered in Refs.~\cite{nashed,biswas2,rej2,vath,sharif,mous,hs}. 

Now to discuss the physical basis of this important stability criteria we 
shall briefly review stellar pulsations. The dynamic equations that govern 
stellar pulsations (Eqs.\ \eqref{eq11}, \eqref{eq12}) can be written in 
several forms \cite{mtw, kokkotas, chanmugam, vath, godenk}. Adapting the form 
given by Misner et al. we may write these equations as \cite{mtw}
\begin{equation}
\label{eq111}
(P\zeta')'+Q\zeta+\omega^2 W\zeta=0
\end{equation}
with
\begin{equation}
r^2 W =(p+\rho)e^{(3\lambda+\chi)/2},
\end{equation}
\begin{equation}
r^2 P=\Gamma\, p\,e^{(\lambda+3\chi)/2},
\end{equation}
\begin{equation}
r^2 Q=e^{(\lambda+3\chi)/2}(p+\rho)\left[(\chi')^2+4\dfrac{\chi'}{r}-8\pi e^\lambda p\right],
\end{equation}
where the Lagrangian displacement has a harmonic time dependence as 
$\zeta=\zeta(r)e^{-i\omega t}$. $\zeta(r)$ and $\omega$ being the amplitude of 
the perturbation and the oscillation frequency, respectively.

We have boundary conditions at the origin $\zeta(r=0)=0$ and at the surface 
$p\,\Gamma\zeta(r)'=0$. In both the relativistic limit and in the Newtonian 
limit, the oscillation problems is a Sturm-Liouville boundary value problem. 
In such cases the same mathematical formulations will work for both 
\cite{kokkotas}. One method for solving such problem is the variational 
technique \cite{bardeen,mtw}. For a Newtonian star the pulsation equation 
\eqref{eq111} reduces to the form:
\begin{equation}
\left[p\,r\,\Gamma(\zeta/r)'\right]'+3\left(p\,\Gamma\zeta/r\right)'-4\,p'\zeta/r+\omega^2\rho\,\zeta=0.
\end{equation}
For such a case the oscillation frequency can be given as
\begin{equation}
\omega^2=(3\overline{\Gamma}-4)|\Omega|/I,
\end{equation}
where $I=\int{(\rho\,r^2)\,4\,\pi r^2 dr}$ is the trace of the second 
moment of the mass distribution, $\Omega$ is the star's self-gravitational 
energy and $\overline{\Gamma}$ being the pressure averaged adiabatic index 
given as
\begin{equation}
\overline{\Gamma}=\dfrac{\int_{0}^{R}p\,\Gamma 4\pi r^2 dr}{\int_{0}^{R}p\,4\pi r^2 dr}.
\end{equation} 
As mentioned earlier for a physical system the necessary and 
sufficient condition for the stability against radial perturbations is that 
the eigenfrequency of the lowest normal mode must be real. From the above 
equation it is clear that $\overline{\Gamma}=\Gamma_{critical}=4/3$ will lead 
to the physically acceptable solution for the fundamental mode of oscillation. 
Thus for $\overline{\Gamma}>4/3$  the Newtonian star is stable in nature and 
will oscillate \cite{mtw}. 
However, in the case of relativistic stars this critical value 
is slightly larger than $4/3$ due to relativistic effects. Hence, a 
sufficiently larger value of $\Gamma$ can ascertain the stability of such 
configurations, which is in fact already clear from the previously shown 
numerically calculated frequency associated to the lowest mode of 
oscillations.

\begin{figure*}[!h]
        \centerline{
        \includegraphics[scale = 0.27]{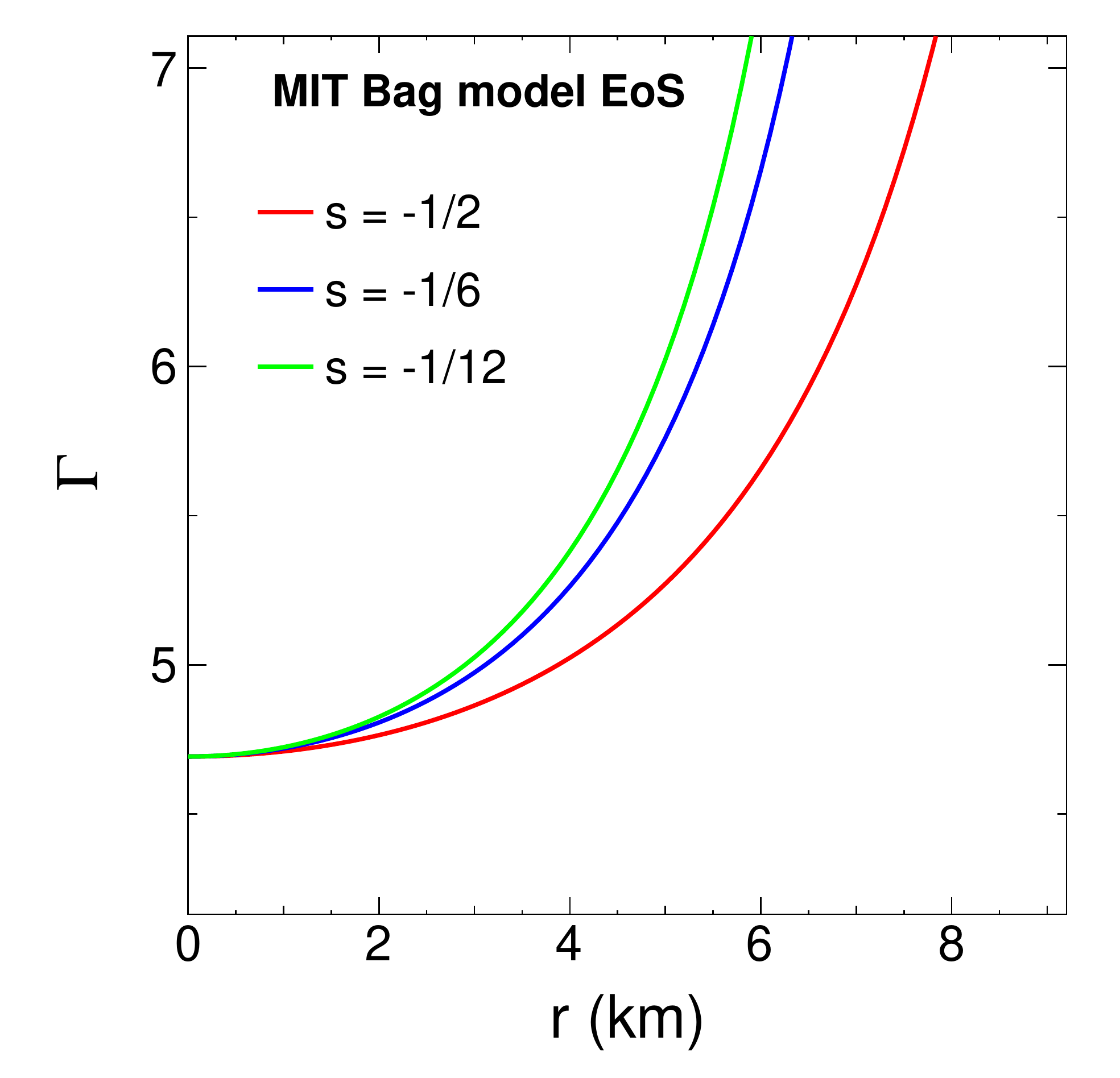}\hspace{0.2cm}
        \includegraphics[scale = 0.27]{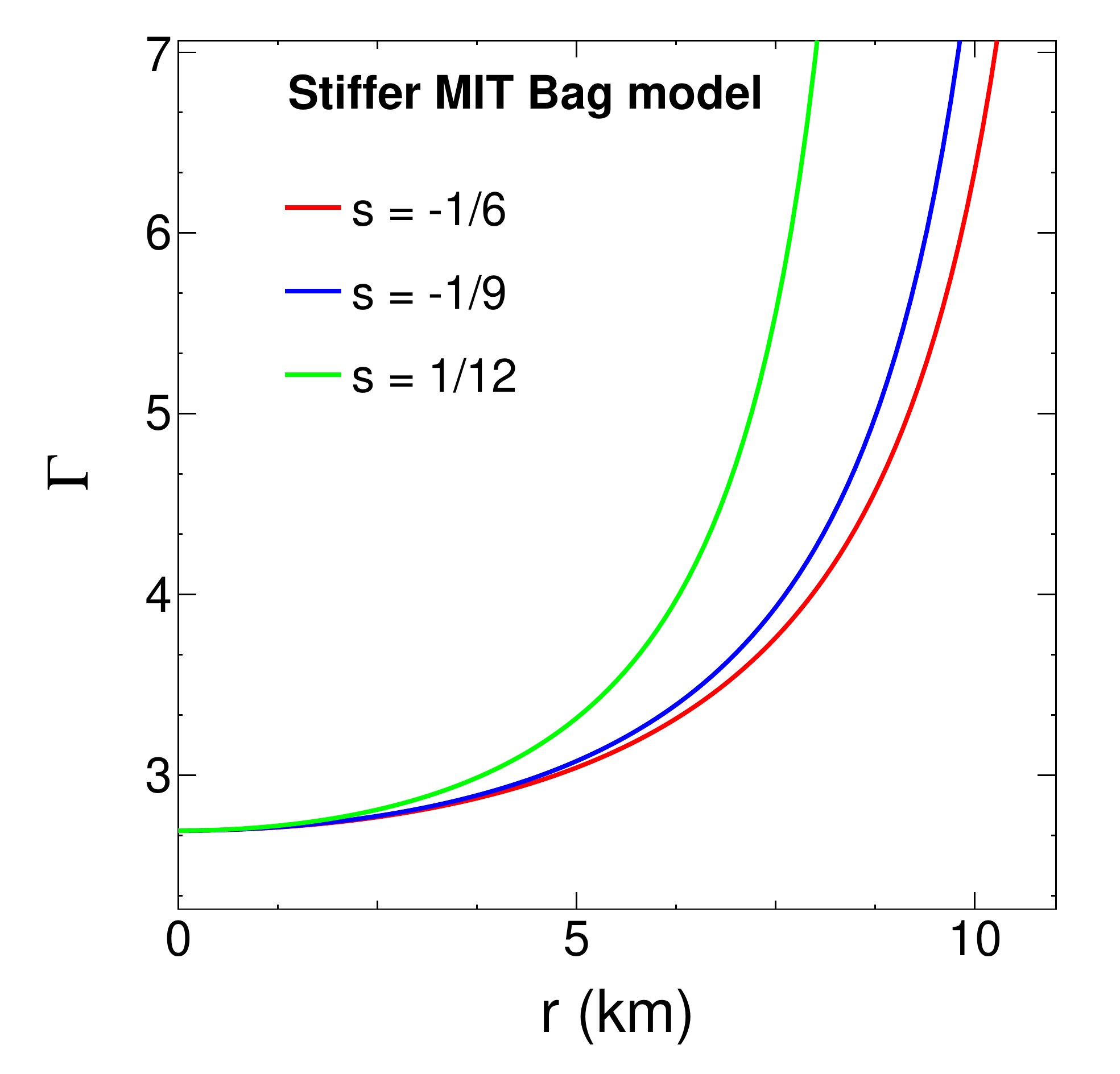}\hspace{0.2cm}
        \includegraphics[scale = 0.27]{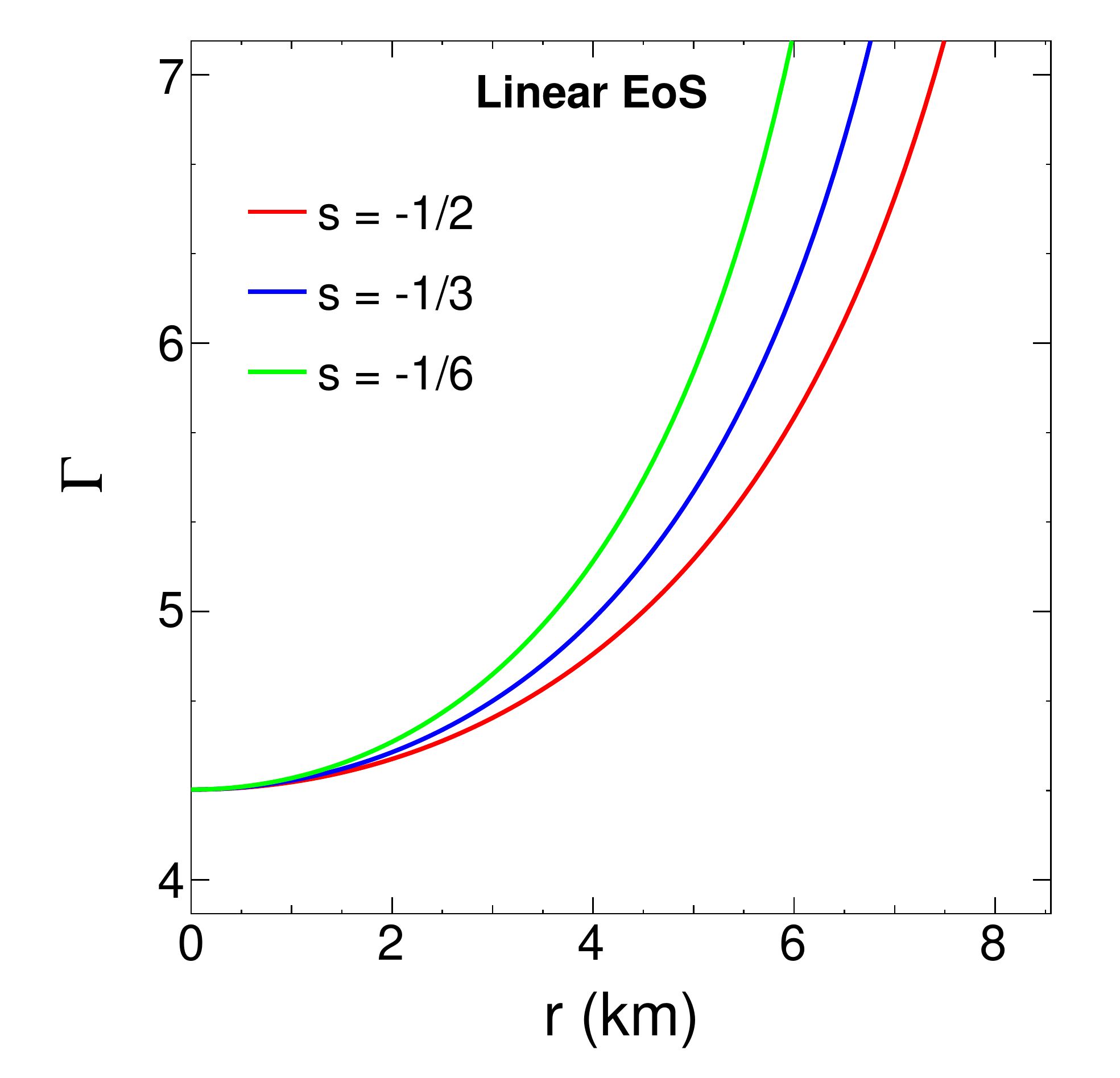}}
        \vspace{-0.3cm}
        \caption{Behaviours of relativistic adiabatic index inside the stellar 
interior in the Starobinsky model for the MIT Bag model EoS (left panel), the
stiffer MIT Bag model EoS (middle panel) and the linear EoS (right panel) 
respectively.}
        \label{fig13}
        \end{figure*}

In Figs.~\ref{fig13}\,-\ref{fig15} the graphical variation of relativistic 
adiabatic index in the stellar interior are shown for three considered 
$f(\mathcal{R})$ gravity models. These variations for the Starobinsky model 
are shown in Fig.~\ref{fig13}. For the Starobinsky model with the MIT Bag 
model EoS the minimum $\Gamma$ value is obtained as $4.69$ 
and the variations of $\Gamma$ for $s=-1/2, -1/6, -1/12$ are shown in the 
first panel of Fig.\ \ref{fig13}. In the second panel of this figure, 
variations of $\Gamma$ are shown for the stiffer MIT Bag model and in the 
third panel the same are shown for the linear EoS. For the stiffer EoS the 
minimum $\Gamma$ value is $2.69$ and for the linear EoS it is $4.33$. As the 
minimum $\Gamma$ value for all these EoSs in the Starobinsky model are well 
above the limiting case, hence they signify the stable configurations.

\begin{figure*}[!h]
        \centerline{
        \includegraphics[scale = 0.27]{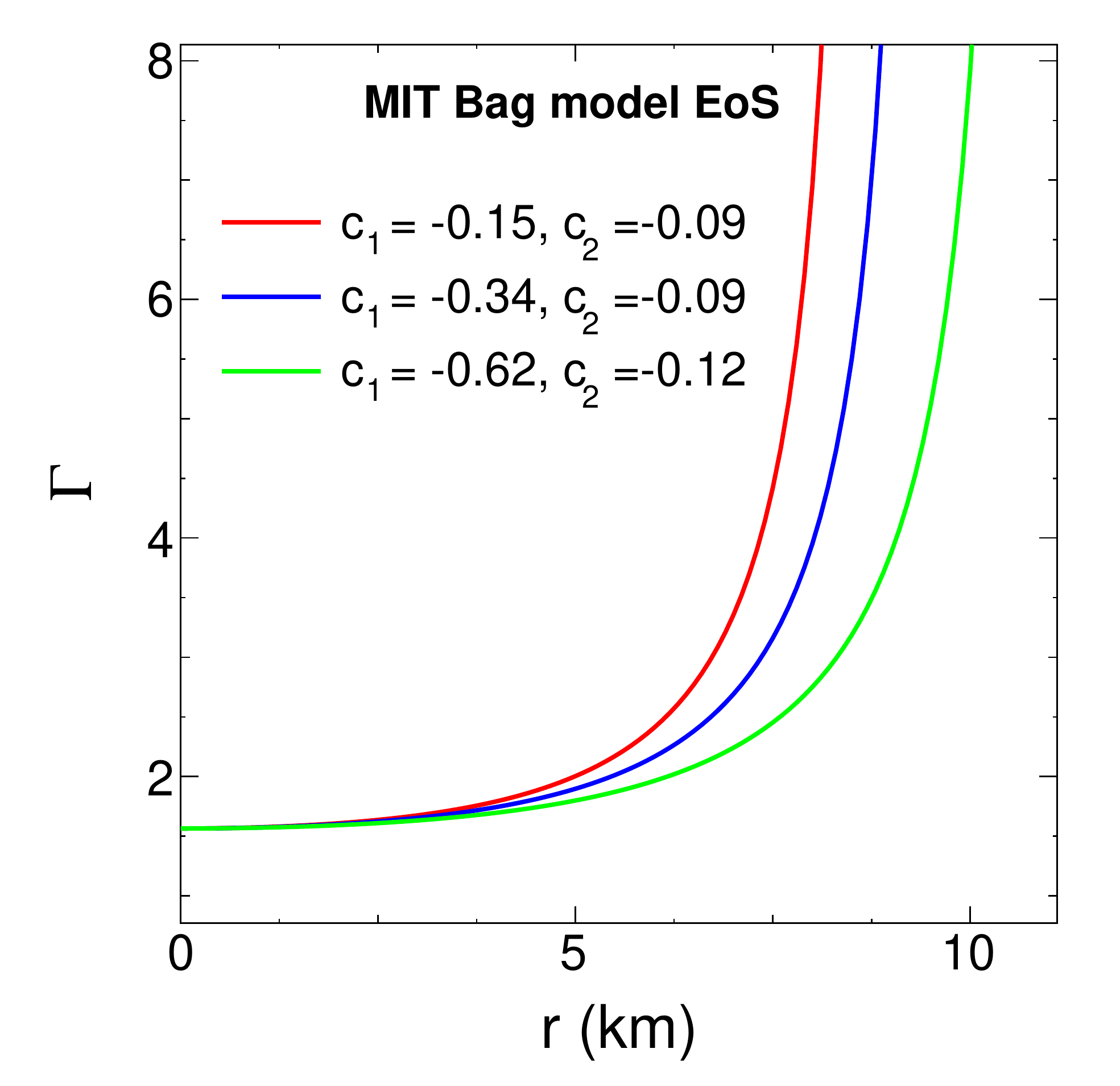}\hspace{0.2cm}
        \includegraphics[scale = 0.27]{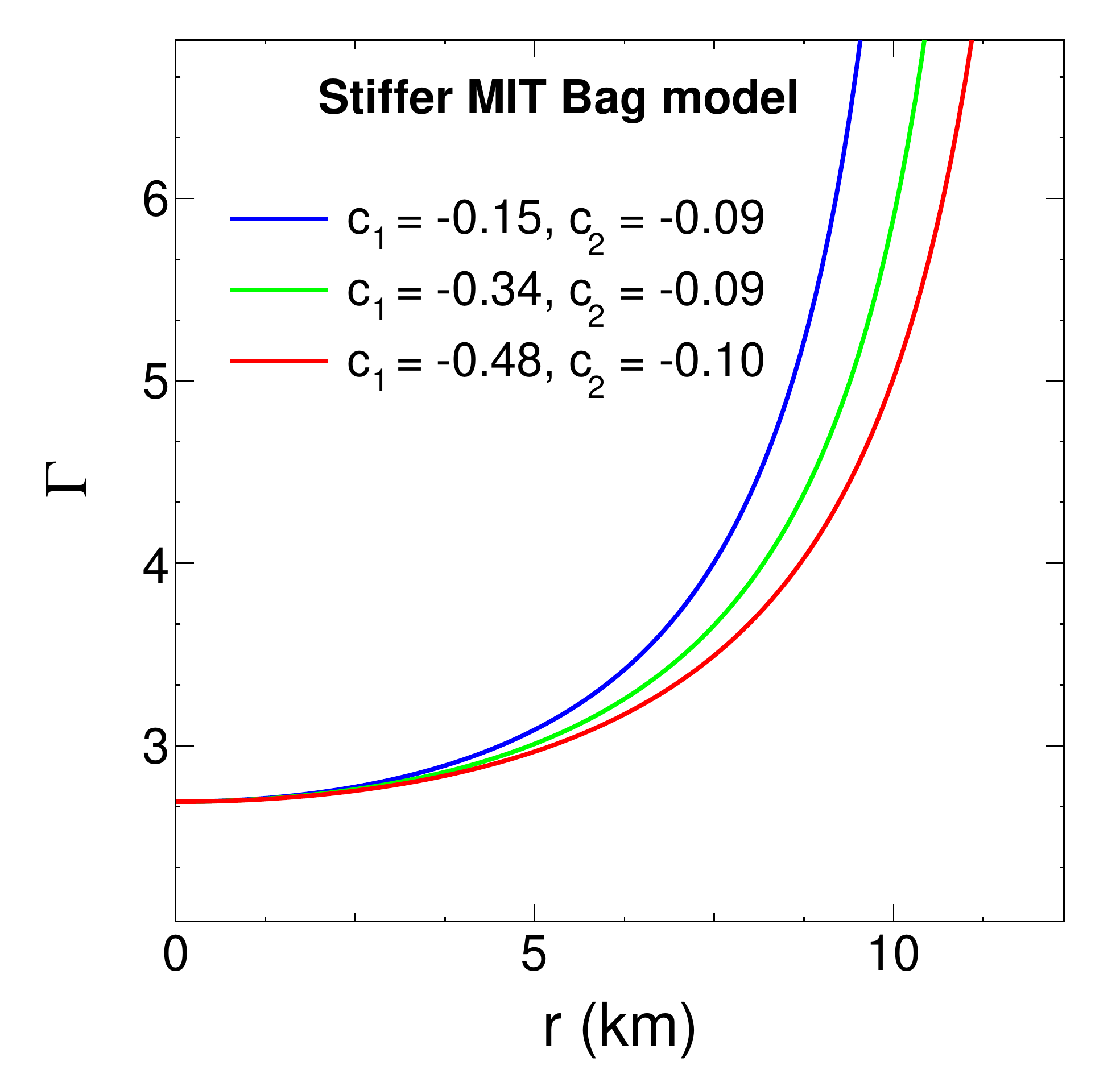}\hspace{0.2cm}
        \includegraphics[scale = 0.27]{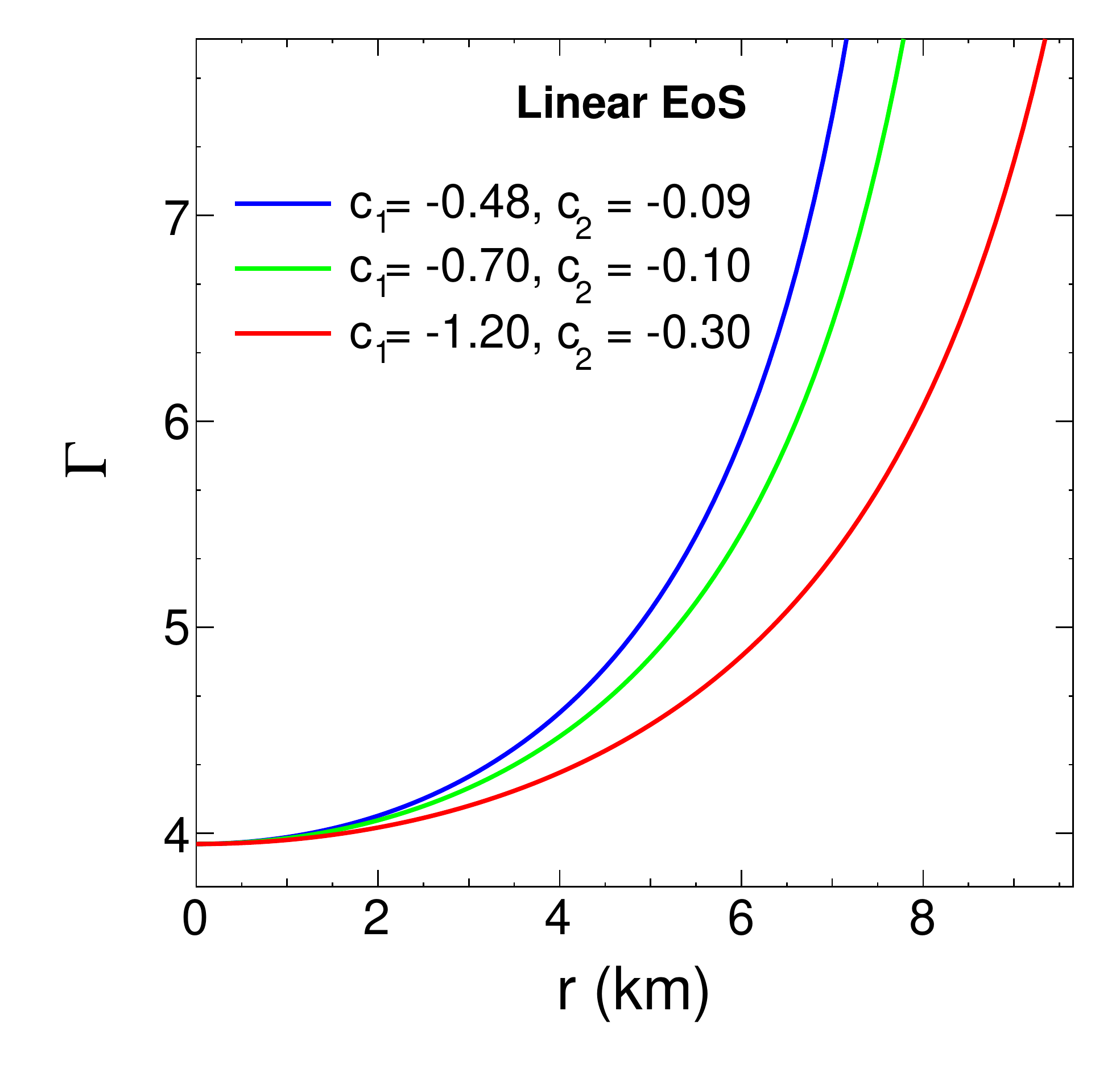}}
        \vspace{-0.3cm}
        \caption{Behaviours of relativistic adiabatic index inside the stellar 
interior in the Hu-Sawicki model for the MIT Bag model EoS (left panel), the 
stiffer MIT Bag model EoS (middle panel) and the linear EoS (right panel) 
respectively.}
        \label{fig14}
        \end{figure*}

\begin{figure*}[!h]
        \centerline{
        \includegraphics[scale = 0.27]{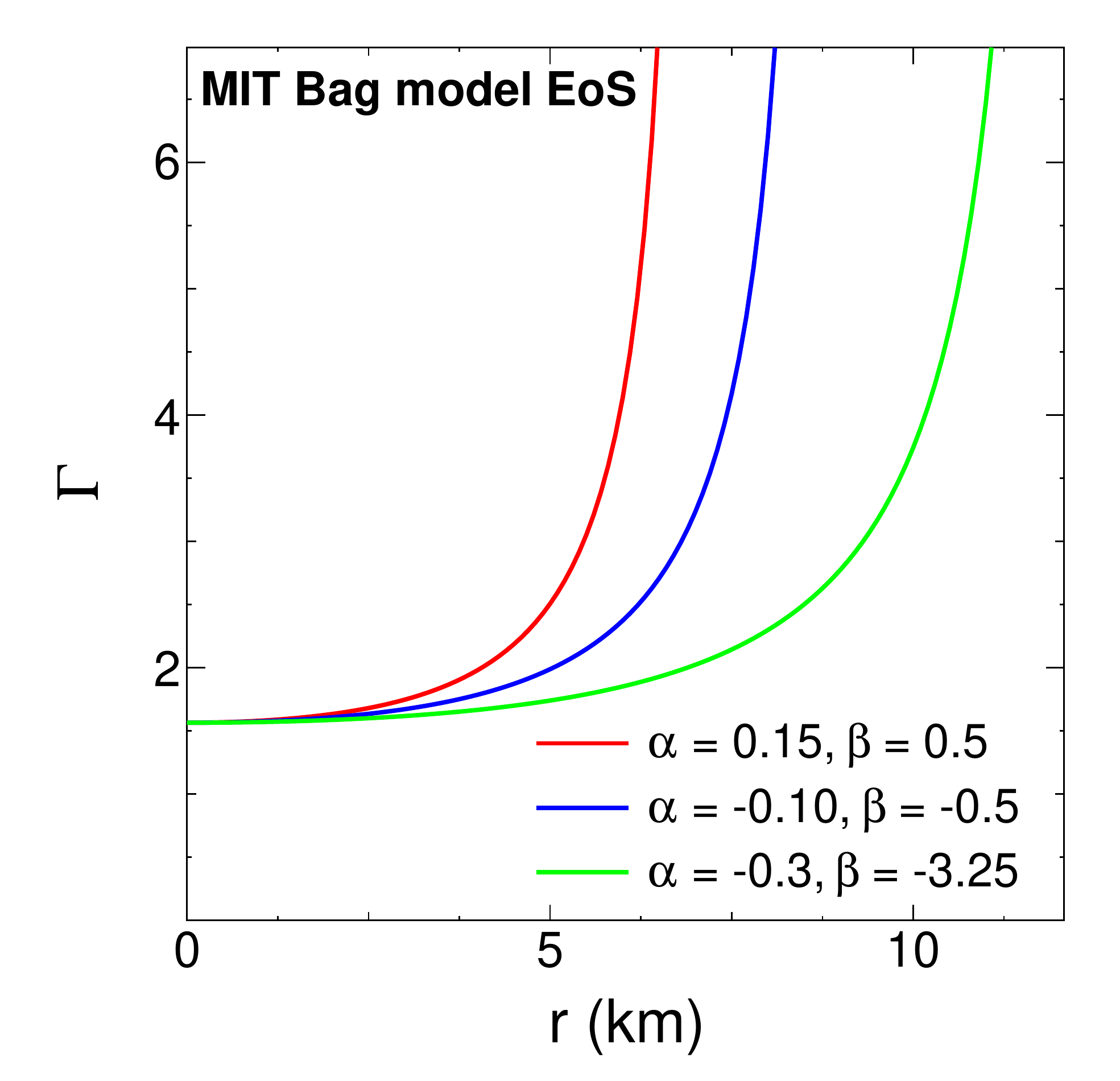}\hspace{0.2cm}
        \includegraphics[scale = 0.27]{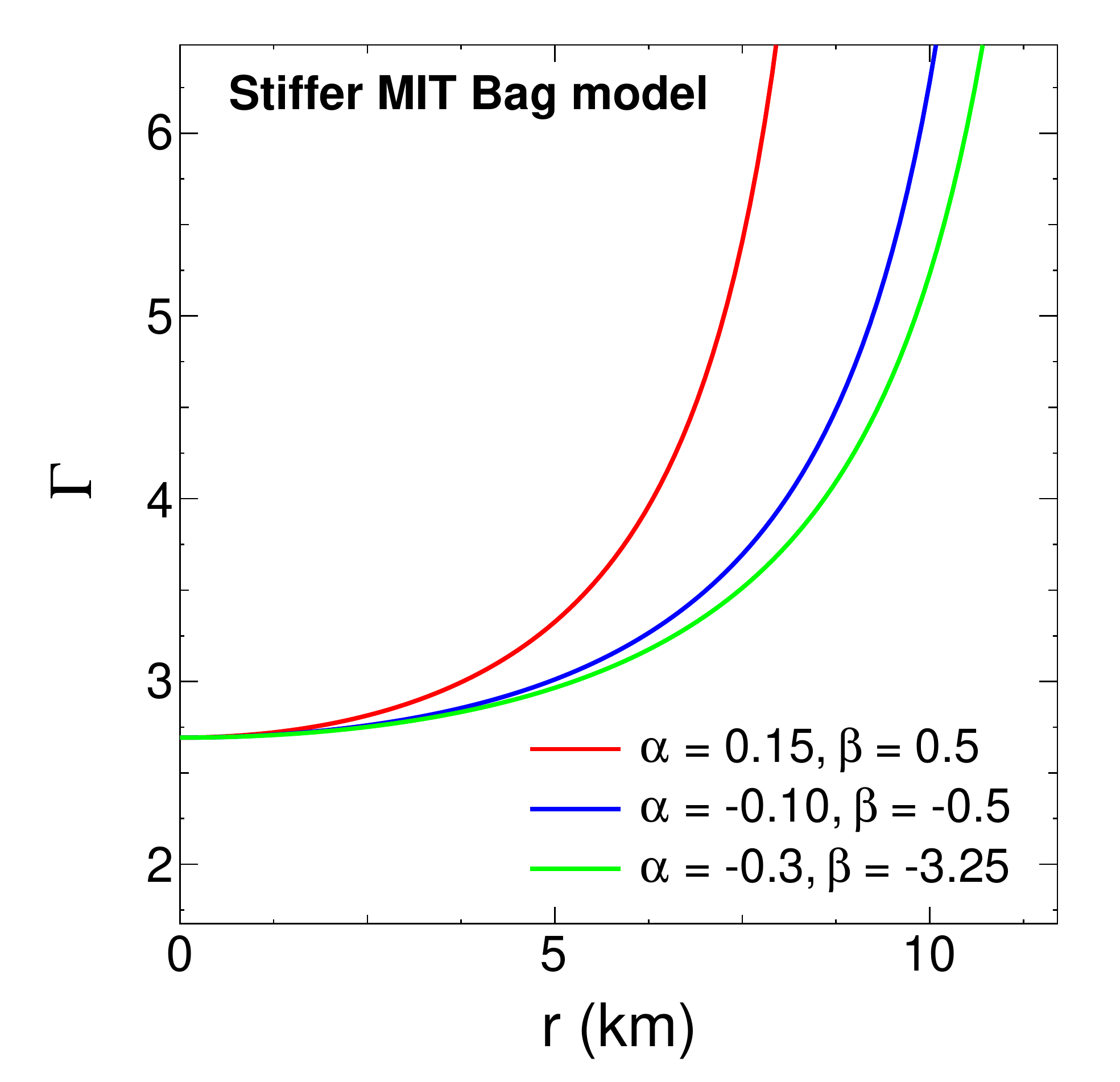}\hspace{0.2cm}
        \includegraphics[scale = 0.27]{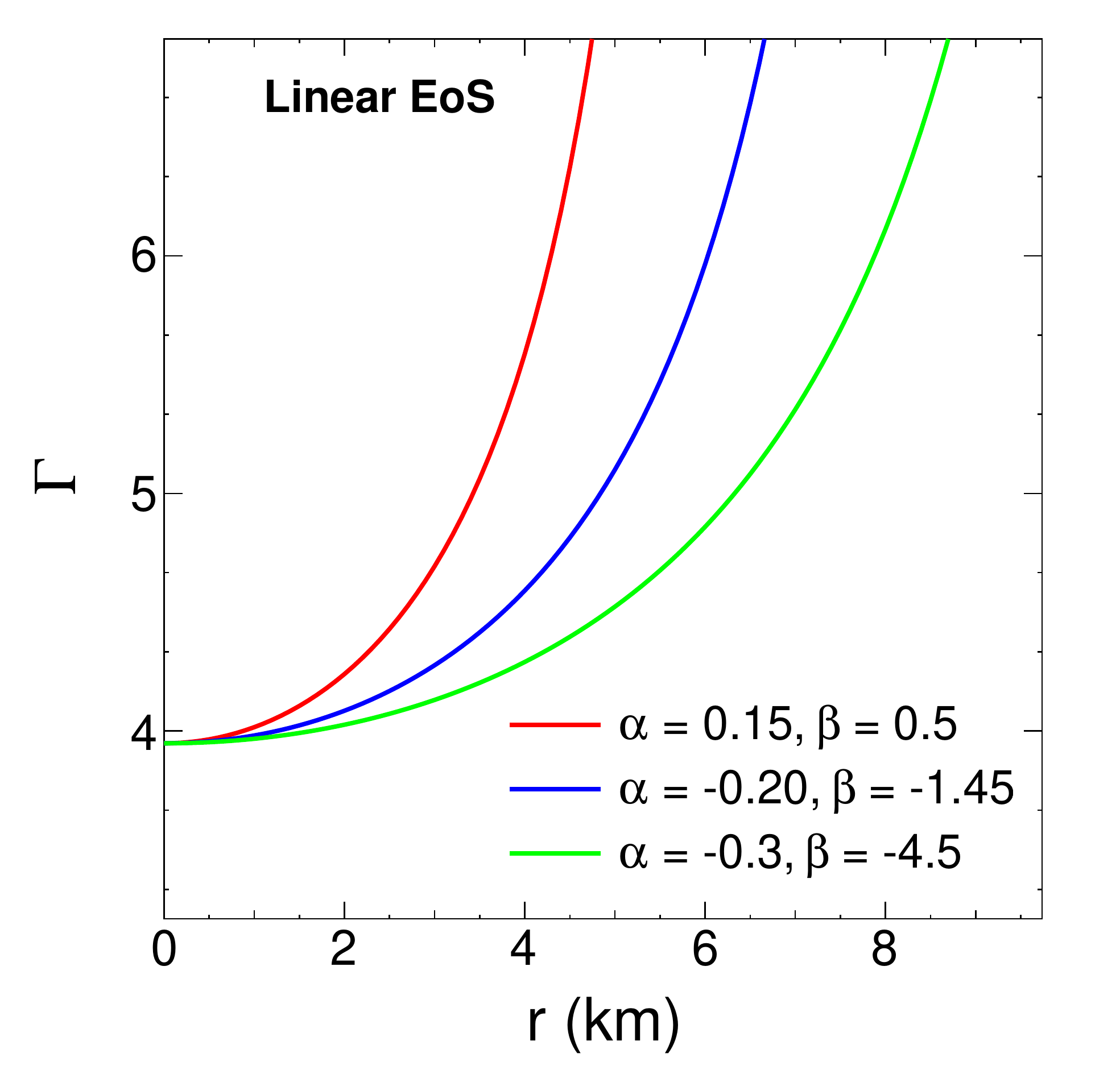}}
        \vspace{-0.3cm}
        \caption{Behaviours of relativistic adiabatic index inside the stellar 
interior in the Gogoi-Goswami model for the MIT Bag model EoS (left panel), the
stiffer MIT Bag model EoS (middle panel) and the linear EoS (right panel) 
respectively.}
        \label{fig15}
        \end{figure*}

Fig.~\ref{fig14} shows the behaviour of $\Gamma$ of strange star configurations
 given by the Hu-Sawicki model for different considered model parameters. The 
central $\Gamma$ value for the stars with the MIT Bag model is $1.56$, for 
the stiffer MIT Bag model it is $2.69$ and for the linear EoS this value is 
$3.95$. Like the case in the Starobinsky model, for this model also solutions 
are stable against the adiabatic index.  

For the case of strange stars predicted by the Gogoi-Goswami model the minimum 
value of $\Gamma$ obtained for the MIT Bag model, the stiffer MIT Bag model 
and the linear EoS are respectively $1.56$, $2.69$ and $3.95$. Variations of
$\Gamma$ are shown in Fig.~\ref{fig15} for three EoSs respectively.
In all the these $f(\mathcal{R})$ gravity models and in considered EoSs, 
$\Gamma$ are taking values well above $4/3$ everywhere inside the stellar 
structure. These figures clearly ensures the non-instability regions inside 
the stellar structures. Fulfilment of this condition shows the stable nature 
of these structures. 

\subsubsection{Surface redshift}
\begin{figure*}[!h]
        \centerline{
        \includegraphics[scale = 0.27]{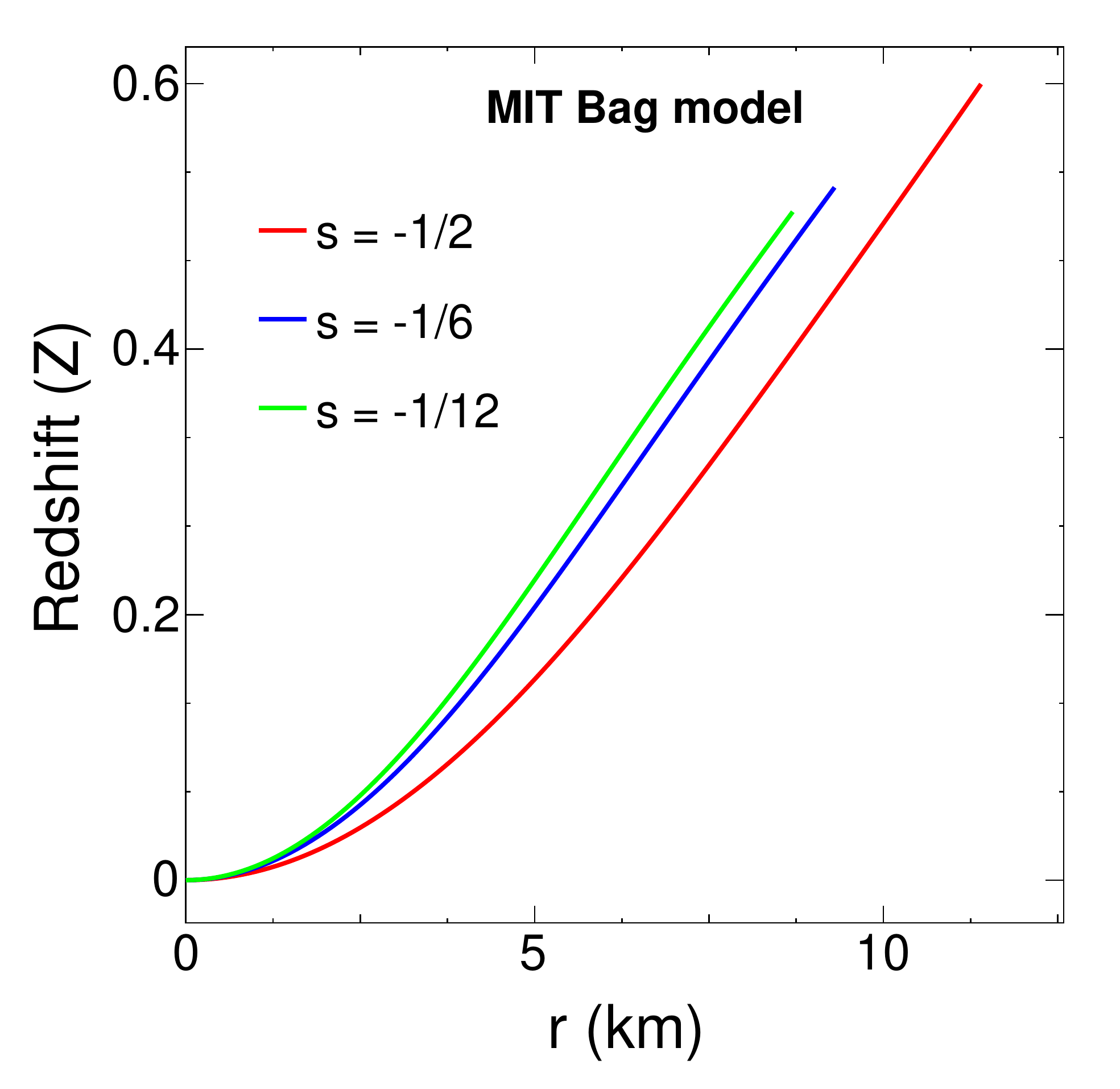}\hspace{0.2cm}
        \includegraphics[scale = 0.27]{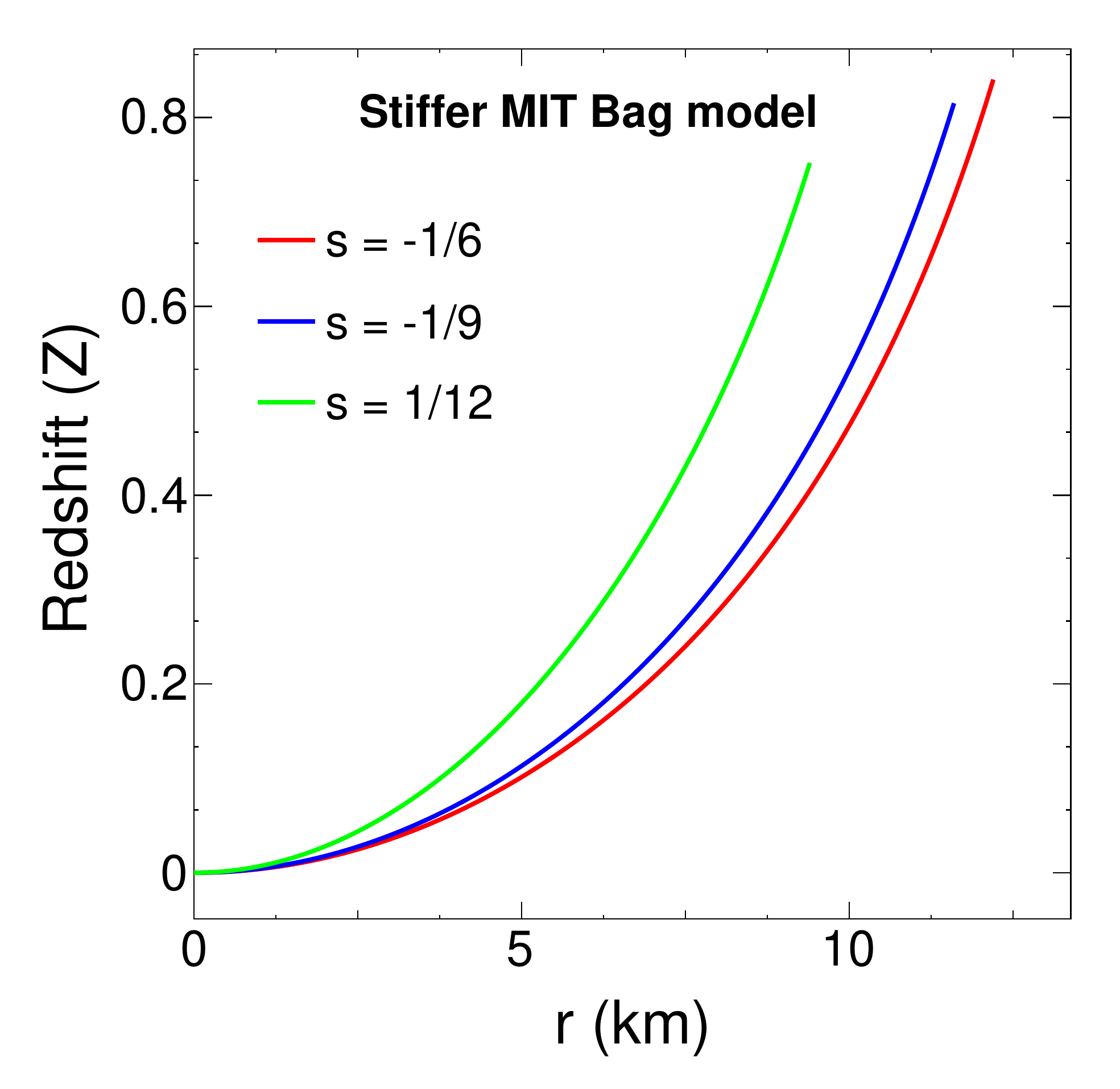}\hspace{0.2cm}
        \includegraphics[scale = 0.27]{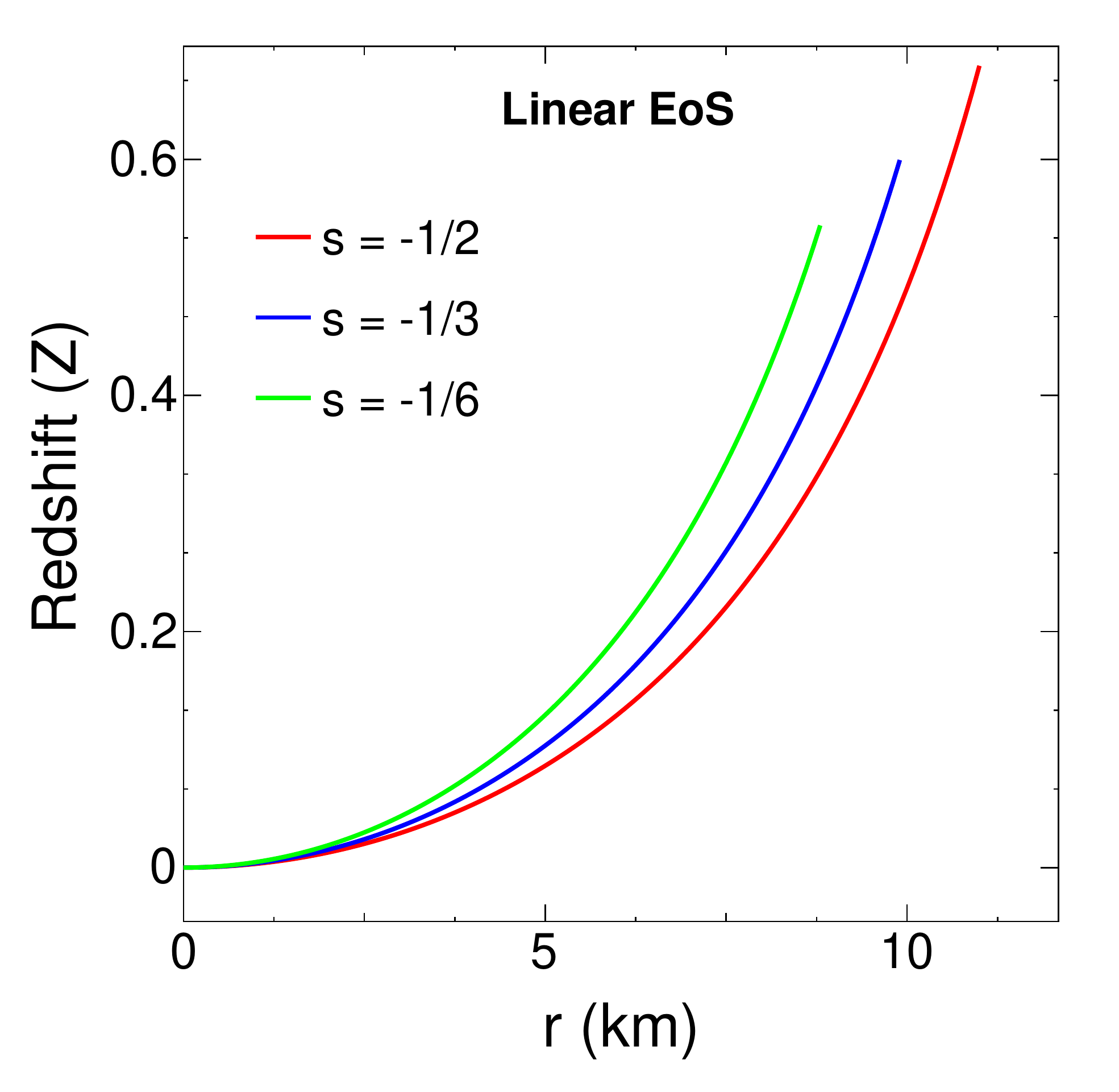}}
        \vspace{-0.3cm}
        \caption{Behaviours of surface redshift with radial distance of strange stars in the Starobinsky model for the MIT Bag model EoS (left panel), the
stiffer MIT Bag model EoS (middle panel) and the linear EoS (right panel)
 respectively.}
        \label{fig16}
        \end{figure*}

\begin{figure*}[!h]
        \centerline{
        \includegraphics[scale = 0.27]{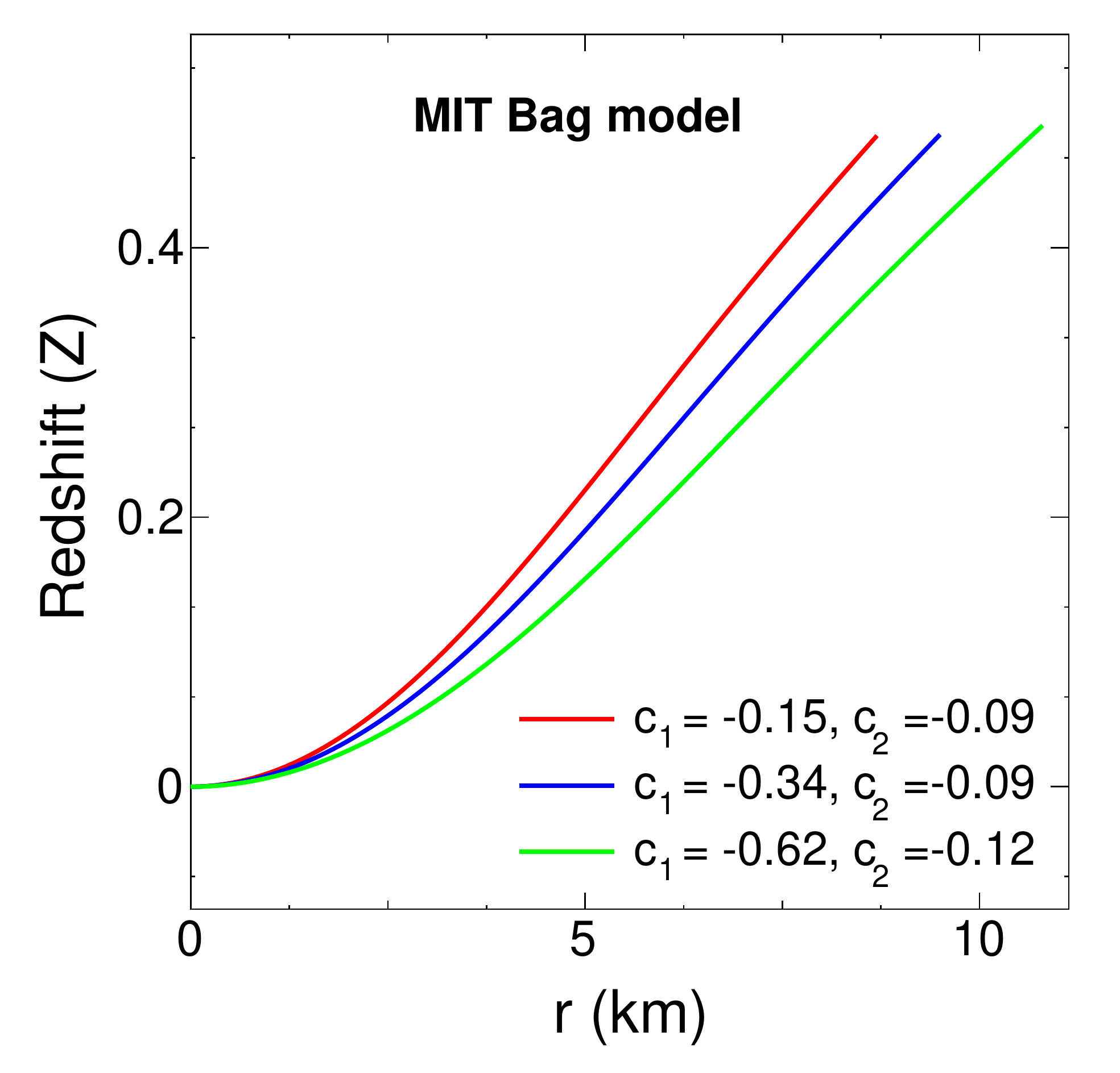}\hspace{0.2cm}
        \includegraphics[scale = 0.27]{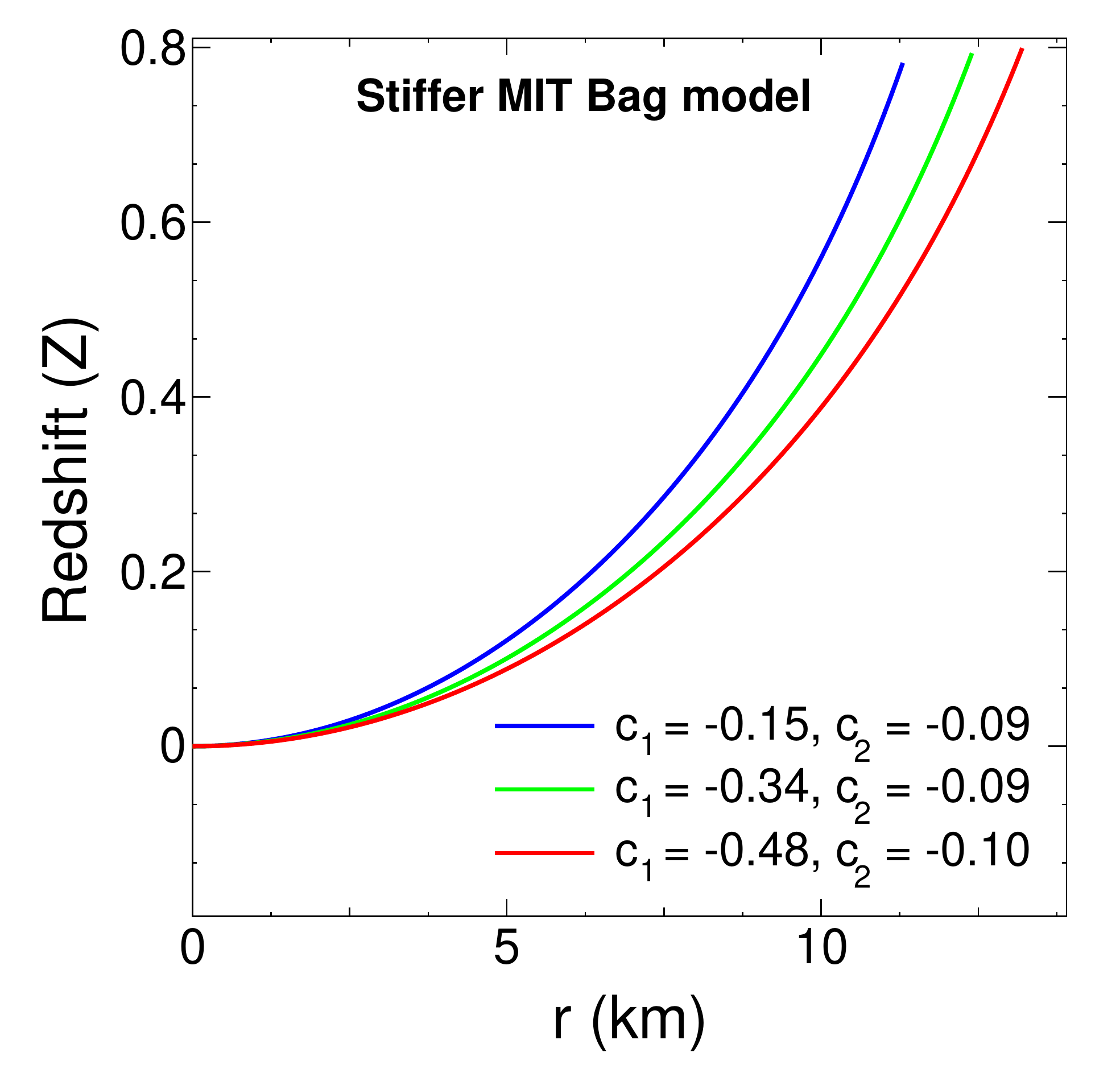}\hspace{0.2cm}
        \includegraphics[scale = 0.27]{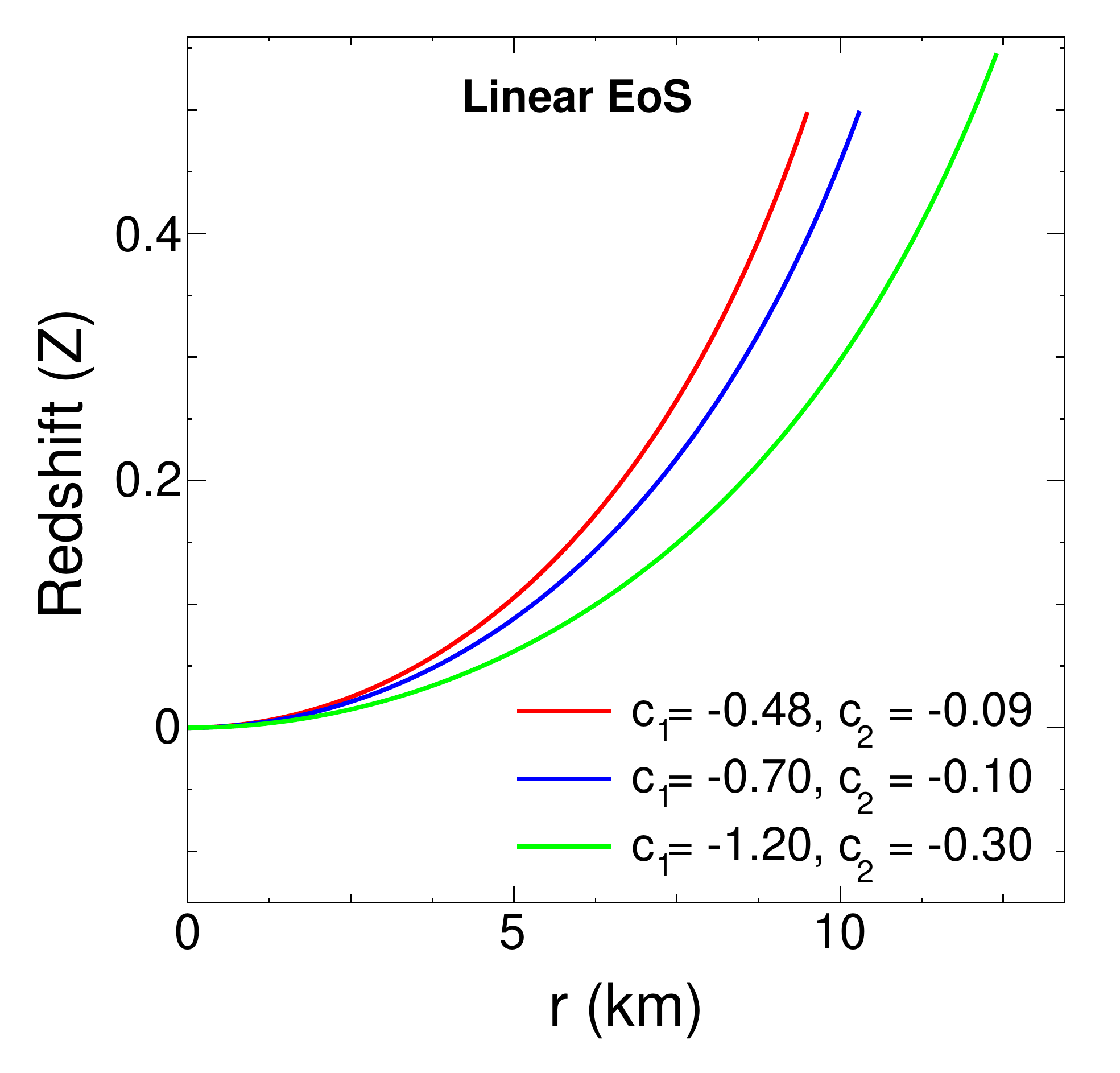}}
        \vspace{-0.3cm}
        \caption{Behaviours of surface redshift with radial distance of strange
stars in the Hu-Sawicki model for the MIT Bag model EoS (left panel), the 
stiffer MIT Bag model EoS (middle panel) and the linear EoS (right panel) 
respectively.}
        \label{fig17}
        \end{figure*}

\begin{figure*}[!h]
        \centerline{
        \includegraphics[scale = 0.27]{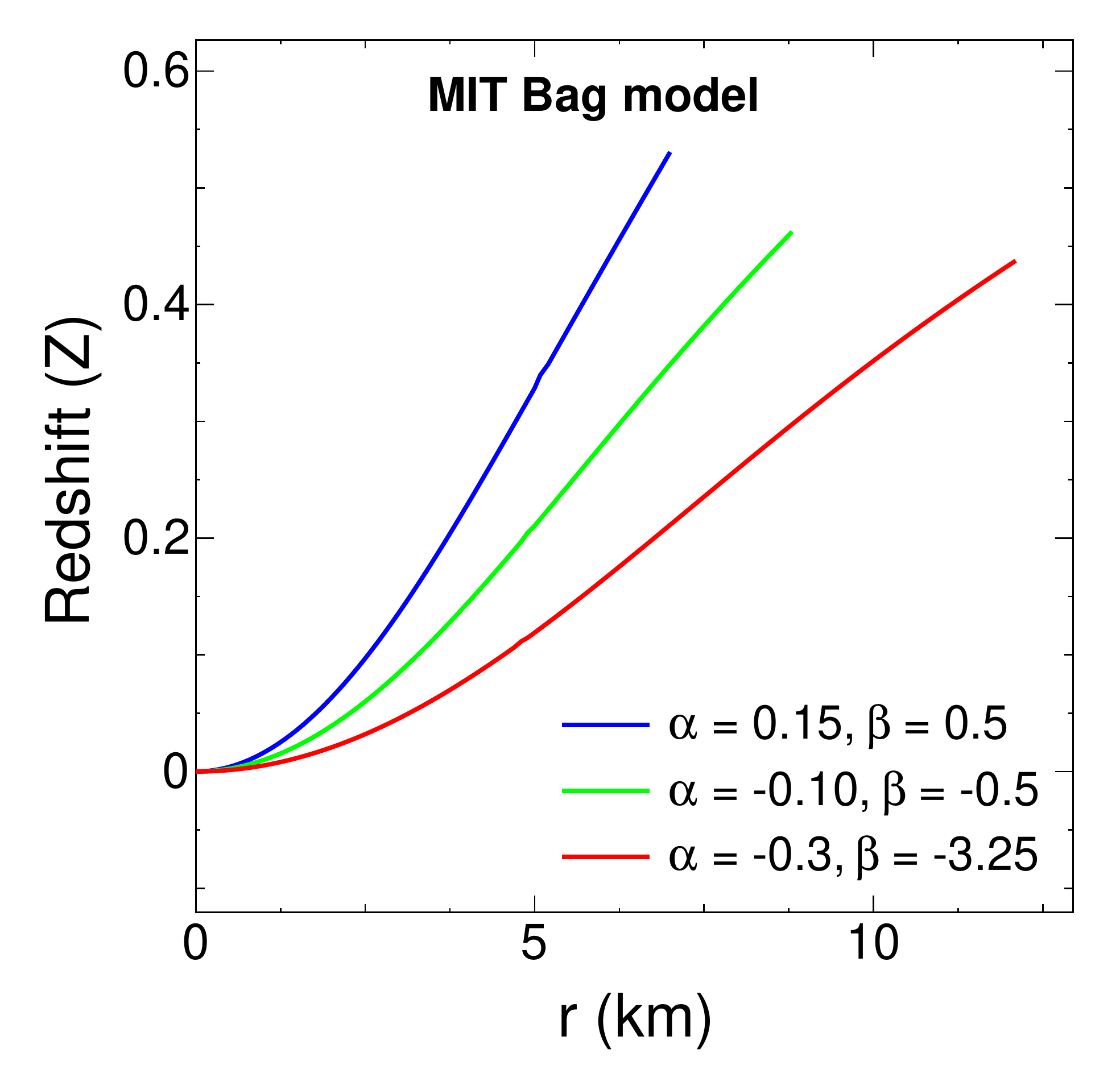}\hspace{0.2cm}
        \includegraphics[scale = 0.27]{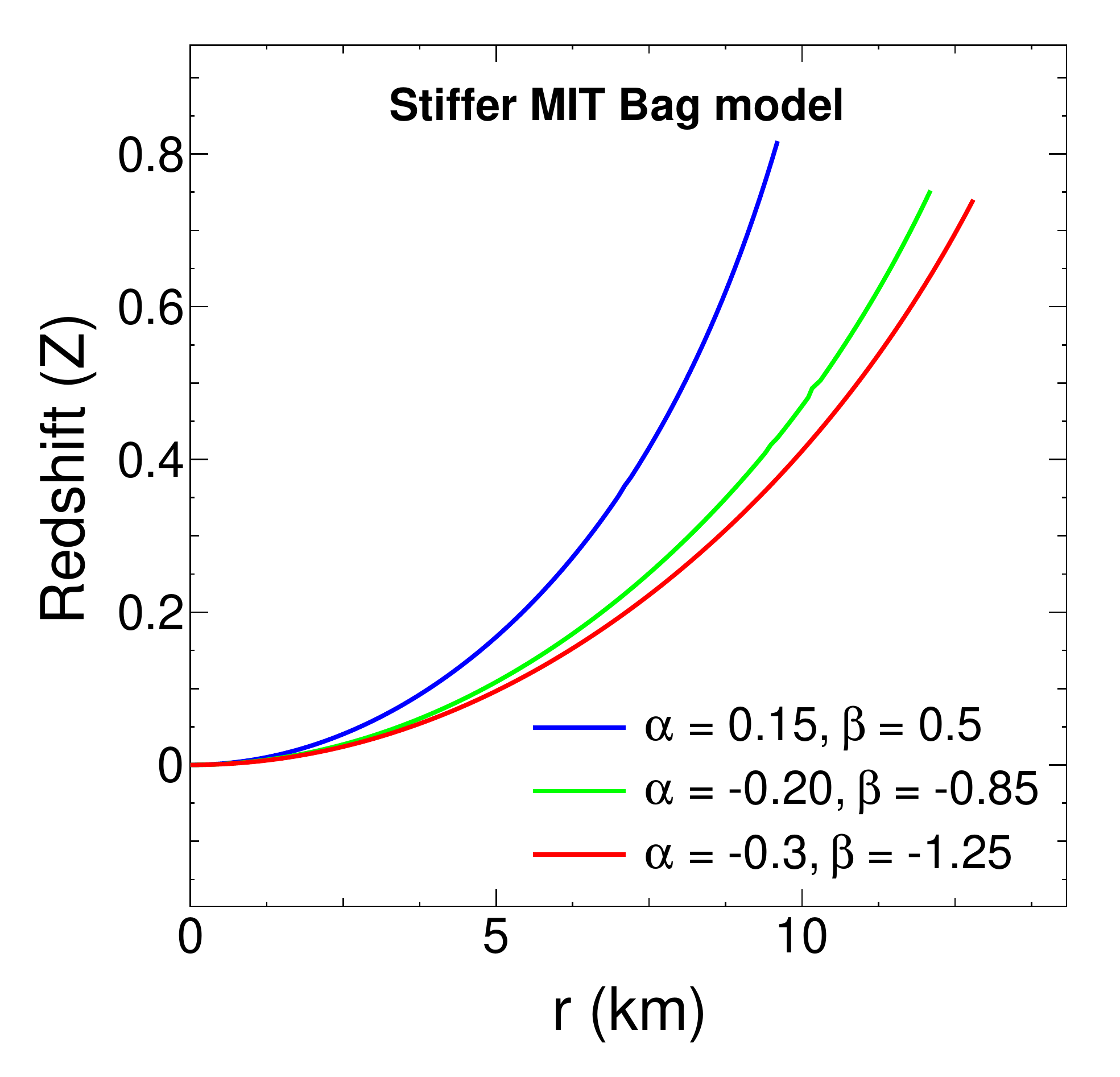}\hspace{0.2cm}
        \includegraphics[scale = 0.27]{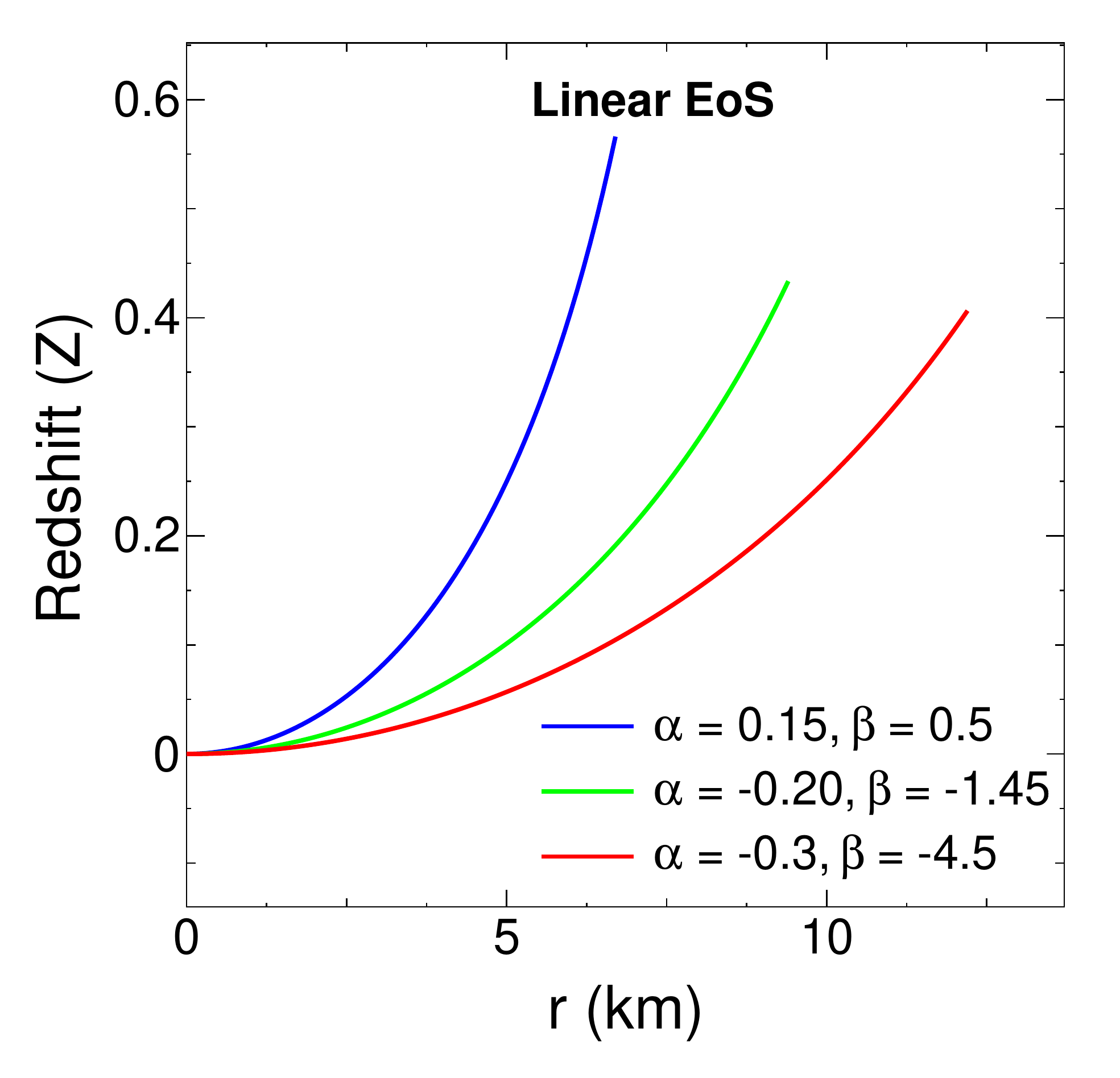}}
        \vspace{-0.3cm}
        \caption{Behaviours of surface redshift with radial distance of 
strange stars in the Gogoi-Goswami model for the MIT Bag model EoS 
(left panel), the stiffer MIT Bag model EoS (middle panel) and the linear EoS 
(right panel) respectively.}
        \label{fig18}
        \end{figure*}

The surface redshift of a star in general describes the relation between the 
interior geometry of the star and its EoS. As mentioned earlier we have 
considered the strange star configuration as isotropic, static, spherically 
symmetric and perfect fluid star. The compactification factor for such a star
 can be given as
\begin{equation}
u(r)=\dfrac{m(r)}{r}.
\end{equation}
Using this compactification factor the surface redshift of a star can be 
defined as
\begin{equation}
Z=\dfrac{1}{\sqrt{(1-2u)}}-1
\end{equation}
For isotropic stars the surface redshift $Z\leq 2$ \cite{buchdahl, baraco}.
The variations of surface redshift with radial distance in the case of 
Starobinsky model for the three EoSs are shown in Fig.\,\ref{fig16}. The 
surface redshift corresponding to this $f(\mathcal{R})$ gravity model is 
listed in Table\,\ref{tab:table2}. For the MIT Bag model EoS the maximum 
surface redshift is $Z=0.600$. For the stiffer EoS, it is $0.840$ and for 
the case of linear EoS, $Z=0.679$. All these obtained redshift values are 
well within the desired range. For the Hu-Sawicki model the $Z$ values 
corresponding to each EoS with the selected sets of model parameter are 
given in Table\,\ref{tab:table4}. Also its variations can be visualized as 
shown in Fig.\,\ref{fig17}. The obtained maximum $Z$ values for this model are 
$0.491$, $0.799$ and $0.546$ for the MIT Bag model, the stiffer MIT Bag 
model and the linear EoS respectively. Again for the Gogoi-Goswami model the 
obtained results are listed in Table\,\ref{tab:table6} and the variations are 
shown in Fig.\,\ref{fig18}. With the MIT Bag model EoS the maximum surface 
redshift is found as $0.531$, for the stiffer MIT Bag model it is $0.817$ 
and for the linear EoS this value is $0.566$. All these values are less than 
$2$ and hence these models predict the stable strange star configurations.
             
%%%%%%%%%%%%%%%%%%%%%%%%%%%%%%%%%%%%%%%%%%%%%%%%%%%%%%%%%%%%%%%%%%%%%%%%%%%%%%%%%%

\section{Summary and Conclusions}\label{conclusion} 
Strange stars in $f(\mathcal{R})$ gravity Palatini formalism have been studied 
in Ref.~\cite{panotopoulos}, where the author used the MIT Bag model EoS to 
study the respective stellar structures. But in this work, we have extended 
the study to three EoSs viz., MIT Bag model, stiffer MIT Bag model and linear 
EoSs for three different viable and well-established dark energy 
$f(\mathcal{R})$ gravity models. It is worth mentioning that strange stars 
were not studied earlier using the later two EoSs in the $f(\mathcal{R})$
Palatini formalism. For the later two EoSs, the trace of the energy-momentum 
tensor becomes the density dependent resulting a complex situation. We have 
solved the corresponding TOV equations numerically and obtained the stellar 
structures for the three different $f(\mathcal{R})$ gravity models using 
different model parameters within their most reliable ranges. Because we have 
chosen the model parameters in compliance with the most promising candidates 
of strange stars obtained so far. The results show that the $f(\mathcal{R})$ 
gravity models support the possibilities of such stable strange stars. 
Moreover, we have also shown that the linear and the stiffer MIT Bag model EoSs
provide the possible strange stars that can echo the GWs. For the Starobinsky 
model with the stiffer EoS, we have seen that the M-R curve for $s=1/12$ shows 
good agreement with most of the observed strange star candidates. For this 
case, the GW echo frequency is found to be $81.829$ kHz with echo time 
$0.038$ ms. On the other hand, for the linear EoS, $s=-1/3$ shows a better 
agreement with the experimentally obtained strange star candidates. The 
corresponding echo frequency is found to be $66.691$ kHz with an echo 
time $0.047$ ms. For the later EoS, we observe a decrease in the echo 
frequency and increase in the echo time. For the Hu-Sawicki model with the stiffer 
EoS, we can see that the model shows a significant deviation from the experimental 
results and the M-R curves are not found in a good agreement with the experimental 
candidates in the GR limit also. On the other hand, for the linear EoS with 
$c_1 = -0.48$ and $c_2 = -0.09$, the M-R curve covers most of the experimental 
candidates with a possibility of GW echo frequency $79.069$ kHz with echo time 
$0.037$ ms. Finally, we considered the Gogoi-Goswami model, in which the stiffer 
MIT Bag model, with $\alpha = 0.12$ and $\beta = 0.34$ shows a better agreement 
with the observed strange star candidates along with a possibility of GW echo 
frequency $76.804$ kHz and echo time $0.041$ ms. While for the linear EoS case,
the M-R curve of $\alpha = -0.20$ and $\beta = -1.45$ shows a good agreement 
with the observed candidates with an echo frequency $84.853$ kHz and 
echo time $0.037$ ms.

These predicted echo frequencies of GWs will get their firm footing once they 
are detected experimentally. The echo frequencies that we have estimated for 
all the considered cases lay above $50$ kHz. So far GWs with 
frequencies of $\sim 20$ Hz - $4$ kHz and with amplitudes of 
$\sim 2\times 10^{-22}$ - $4 \times 10^{-24}$ strain/$\sqrt{\mbox{Hz}}$ \cite{martynov, abbot} are projected at GW detectors like, Advanced LIGO \cite{ligo}, Advanced Virgo \cite{virgo} and KAGRA \cite{kagra}. Currently 
running LIGO \cite{currentligo} and Virgo \cite{currentvirgo} observatories have a 
sensitivity of $\ge 2\times 10^{-23}$ strain/$\sqrt{\mbox{Hz}}$ at $3$ kHz. The 
third generation detectors, such as Cosmic Explorer (CE) \cite{abbott} and 
Einstein Telescope (ET) \cite{punturo} with optimal arm length of $\approx 20$ km 
would have the sensitivity to detect the amount of postmerger neutron star 
oscillations \cite{punturo}. The sensitivity of CE may reach below $10^{-25}$ 
strain/$\sqrt{\mbox{Hz}}$ at above few kHz frequencies and it is a proposed $40$ 
km arm length L-shaped observatory. On the other hand, ET - the L-shaped 
underground proposed observatory will be able to reach the sensitivity of 
$> 3\times10^{-25}$ strain/$\sqrt{\mbox{Hz}}$ at $100$ Hz and of 
$\sim 6\times10^{-24}$ strain/$\sqrt{\mbox{Hz}}$ at $\sim 10$ kHz. 
Thus, the present and near future GW detectors are not enough sensitive to 
detect such weak signals of GWs. However, the enhancement of the sensitivity
of these detectors are also possible. It is proposed that the sensitivity of 
these detectors can be enhanced by an optical configuration of detectors using 
the current interferometer topology to reach $\ge 7\times 10^{-25}$ 
strain/$\sqrt{\mbox{Hz}}$ at $2.5$ kHz \cite{martynov}. Also, S.\ L.\ 
Danilishin et al.\ \cite{danilishin} proposed that by the application of advanced 
quantum techniques to suppress the quantum noise at high frequency end in 
the design of GW detectors, the sensitivity of the present GW detectors can be 
enhanced significantly. If applications of such techniques or other possible
method can enhance the sensitivity of GW detectors, it will open a new door 
to explore such echoes of the GWs. Which will definitely throw more light on 
the mystery of stellar interior of compact stars.

An important point to be noted here is that the $f(\mathcal{R})$ gravity 
models considered in this study are the dark energy $f(\mathcal{R})$ gravity 
models, which can mimic the $\Lambda$CDM model at high curvature regime and GR 
in the low curvature regime. The general Starobinsky model also behaves in the 
similar manner but the special case we have considered in this work is the 
inflationary Starobinsky model which is well known for its capabilities of 
explaining the theoretical inflationary epoch of the universe. On the other 
hand, the Gogoi-Goswami model and the Hu-Sawicki model show very promising 
behaviour in the cosmological perspectives and both the models, in Palatini 
formalism can explain the present universe and far past scenario of the 
universe effectively \cite{ijmpd}. At far past or very high redshift range both 
the models are identical to the $\Lambda$CDM model, which makes it almost 
difficult to differentiate the models in high curvature regime. But our studies 
showed that the compact star structure at very high curvature has a high model 
dependency which provides us a new avenue in constraining the models in high 
curvature regime. So, undoubtedly studies of such modified gravity models in 
the area of compact stars will provide a better constraint on the model 
parameters in the high curvature regime once we have a good number of 
experimental results in this direction.

From these results, we can conclude that the $f(\mathcal{R})$ gravity in 
Palatini formalism allows the formation of stable strange stars, and can 
explain the stability and existence of the experimentally obtained strange 
star candidates. Nevertheless, to have a clear picture of the stability 
of such stellar configurations, a more detailed analysis throughout the 
complete 
model parameter space would be necessary. As this sort of study would be 
beyond the scope of the present one, so we leave this as a future prospect 
of the study. We have also provided the echo frequencies and times for such 
compact structures. In the near future, with the experimental data of GW 
echoes from such candidates it might be possible to select the most promising 
EoS or constraining the EoSs, which will provide a better understanding
of such configurations and hence will help to reveal the actual properties of 
such stars. Consequently, the physics of near and the high regime will be 
solved clearly afterwards.

%%%%%%%%%%%%%%%%%%%%%%%%%%%%%%%%%%%%%%%%%%%%%%%%%%%%%%%%%%%%%%%%%%%%%%%%%%%%%%%%%%

\section*{Acknowledgements}
Authors are thankful to the esteemed anonymous referee for his/her useful suggestions which helped us to make 
significant improvements in the manuscript. UDG is thankful to the 
Inter-University Centre for Astronomy and Astrophysics (IUCAA), Pune, 
India for the Visiting Associateship of the institute.

\appendix*

\bibliographystyle{apsrev}
\end{document}